\documentclass[cits,hyper]{JINST}

\usepackage{amsmath}

\usepackage{amsthm}
\newtheorem{prop}{Property}

\usepackage[squaren]{SIunits}

\usepackage{rotating}

\newcommand{\di}{\ensuremath{\,\text{d}}}         

\newcommand{\doi}[1]{\href{http://dx.doi.org/#1}{doi: #1}}
\newcommand{\arxiv}[1]{\href{http://arXiv.org/abs/#1}{arXiv: #1}}

\title{Reference analysis of the signal + background model in counting
  experiments}

\author{Diego Casadei\\
  New York University, 
  4 Washington Place, NY-10003 New York, USA
  \\
  \email{diego.casadei@cern.ch}
}

\abstract{
  The model representing two independent Poisson processes, labelled
  as ``signal'' and ``background'' and both contributing additively to
  the total number of counted events, is considered from a Bayesian
  point of view.  This is a widely used model for the searches of rare
  or exotic events in presence of a background source, as for example
  in the searches performed by high-energy physics experiments.  In
  the assumption of prior knowledge about the background yield, a
  reference prior is obtained for the signal alone and its properties
  are studied.  Finally, the properties of the full solution, the
  marginal reference posterior, are illustrated with few examples.
}


\begin{document}

\maketitle


\section{Introduction}

 Searches for rare events or faint signals are very common in
 scientific research.  Here we consider only counting experiments, as
 for example underground detectors which measure the result of
 high-energy particle collisions, and we are interested in situations
 in which a known set of ``background'' processes contributes to the
 number of observed events, on top of which one looks for a possible
 excess which can be attributed to the faint signal.  In these
 situations, most searches do not find evidences for a new signal and
 their results are summarized by providing upper limits to the signal
 intensity (see for example \cite{casadeiPHSTAT2011}).  The analysis
 of the significance of an experimental result or of the expected
 significance of a planned measurement is a very important task, and
 the problem has been treated several times in the frequentist
 framework (see for example \cite{bityukov,ccgv2011}).  Here the
 problem is approached from the Bayesian point of view, as it has been
 recently done by Demortier, Jain, and Prosper in
 Ref.~\cite{demortier2010} (hereafter named the ``DJP paper''), whose
 work stimulated the present study and a similar paper by Pierini,
 Prosper, Sekmen and Spiropulu al.~\cite{pierini2011} (hereafter named
 the ``PPSS paper'')\footnote{The PPSS paper appeared on arXiv few
 days before this paper, in which it has been referenced during the
 review process.  They report about independent developments on a
 similar subject.}

 We assume that the integer number $k=0,1,\ldots$ of observed events
 follows a Poisson distribution and that the signal and background
 sources are independently Poisson distributed with parameters $s$ and
 $b$, such that the probability to observe $k$ counts comes from a
 Poisson distribution with parameter $s+b$: $P(k) =
 \text{Poi}(k|s+b)$.  Here we are interested in the signal intensity
 $s\ge0$, as it is practically always the case in the searches for new
 phenomena, and treat $b\ge0$ as a nuisance parameter.

 In the Bayesian framework, the result of the statistical inference is
 provided by the joint posterior probability density
\begin{equation}\label{eq-bayes-theorem}
  p(s,b|k) \propto \text{Poi}(k|s+b) \, p(s,b)
\end{equation}
 where $p(s,b)$ is the prior density which encodes the experimenter's
 degree of belief before incorporating the results of the experiment
 with the information available before performing the experiment, and
 the likelihood function is the Poisson model $\text{Poi}(k|s+b)$
 itself, considered as a function of $s+b$ for $k$ fixed at the actual
 observation.  The normalization constant can be found by imposing
 that the integral of $p(s,b|k)$ is one.

 Here we look for a solution which only depends on the assumed model
 and the observed data, but not on any additional prior information,
 hence choose to follow the approach dictated by the so-called
 \emph{reference analysis} \cite{Bernardo2005a}.  A key ingredient is
 the formulation of the \emph{reference prior} $\pi(s,b)$
 \cite{Bernardo2009a}, defined as to maximize the amount of missing
 information, and the result of the inference is the \emph{reference
 posterior} obtained by using $\pi(s,b)$ in place of $p(s,b)$ in the
 Bayes' theorem~(\ref{eq-bayes-theorem}).  Because we are not interested
 in making inference about $b$, we integrate the posterior over it
 to obtain the marginal posterior $p(s|k) = \int_{0}^{\infty}
 p(s,b|k) \di b$, our final solution.  This approach has been followed
 also by the DJP and PPSS papers.

 Although DJP considered signal and background as independent Poisson
 processes as in this work, they preferred a model in which the signal
 strength is multiplied by a parameter which should encode the
 uncertainty about both the signal efficiency and the integrated
 luminosity.\footnote{This means that their signal and background
   strengths represent the cross-sections of the corresponding
   processes, whereas in this paper the two parameters represents the
   expected counts.}  However, the background parameter is not
 multiplied by a similar quantity, although one expects to have some
 luminosity dependent factor here too.  A better model would describe
 the probability of counting $n$ events as the Poisson probability
\(
  \text{Poi}(n|(\varepsilon_s \sigma + \varepsilon_b \mu)\mathcal{L})
\)
 in which the signal cross-section $\sigma$ is multiplied by the
 signal efficiency $\varepsilon_s$, the background cross-section $\mu$
 is multiplied by the corresponding efficiency $\varepsilon_b$, and
 the integrated luminosity $\mathcal{L}$ explicitly appears, because
 it affects the same way signal and background yields (being 100\%
 correlated).  Such model would be quite complex: $\sigma$ is the only
 parameter of interest, hence the marginal posterior is obtained after
 the integration over the remaining four nuisance parameters.  The
 prior for $\mu$ would be a Gamma density as in the DJP paper (the
 Gamma density is the conjugate prior for the Poisson process), the
 priors for the efficiencies $\varepsilon_s , \varepsilon_b$ would be
 Beta densities (the conjugate family for the binomial problem), and
 the prior for $\mathcal{L}$ would be a Gaussian (or a Gamma density
 peaked very far from zero, looking very similar to a Gaussian).  For
 the purpose of the present work, this model is too complicated.  To
 simplify things, we treat a model in which there are only two
 parameters $s,b$ describing the expected numbers of events coming
 from the signal and background processes, assumed to be mutually
 independent (and independent from other parameters which do not
 appear explicitly in the model, like the luminosity, as it is done in
 the PPSS paper).

 Here $s$ and $b$ are dimensionless numbers (not cross-sections) which
 are sufficient to describe in a stylized way the ``discovery
 problem'' in which one is interested in finding an excess of events
 over the background prediction.  Qualitatively, in case a
 statistically significant excess is found, the next step is to
 evaluate the signal cross-section which corresponds to the
 measurement, which requires adopting the complicated model above,
 with a single parameter of interest and four nuisance parameters.
 Although in practice one works since the very beginning with a very
 complex model (with several parameters modelling the detector
 response and the theoretical uncertainties), from the mathematical
 point of view discovery and cross-section measurements can be
 considered two distinct phases, which can be treated in sequence.
 Here we consider the statistical approach which is sufficient for the
 discovery problem in a counting experiment for which the total
 uncertainty on the background yield is the only nuisance parameter,
 and defer the treatment of the complex model which is needed for
 estimating the cross-section to another paper.

 In this paper we do not address in details the general Bayesian
 approach to discovery, but focus only on the marginal posterior for
 the signal in presence of backgrond with unknown yield (which is a
 necessary but not sufficient ingredient).  However, the Reader should
 be aware that there are several issues in the use of the ``Bayes'
 factor'' when taking a decision among two or more possible hypotheses
 when improper priors are used, as recently summarized by Berger
 \cite{Berger2011}.  The widespread use of flat priors in Bayesian
 computations basically makes it impossible to take a decision based
 on the Bayes factor in a proper way, and this is true also for the
 reference prior in the Poisson model considered here (see below),
 which is improper.  Instead, the two posterior solutions which
 correspond to the competing hypotheses should be compared.  This
 approach has no hidden trap when improper priors are chosen, and the
 only difficulty is that there is no consensus about the threshold
 which the posterior ratio should exceed to claim a discovery,
 contrasting with the conventional frequentist ``five-sigma'' rule for
 the significance of the discrepancy with respect to the
 background-only hypothesis (which is not free of problems
 \cite{Demortier2011}).  A promising formal approach based on the
 Bayesian decision theory is being proposed by Bernardo
 \cite{Bernardo2011}, which is based on decision theory.  One would
 choose the hypothesis which minimizes the expected posterior loss
 incurring when acting as if that hypothesis were correct.  A
 convenient choice of a loss function which is parametrization
 invariant is the so-called ``intrinsic discrepancy loss'', defined as
 the minimum among the two Kullback-Leibler directed divergences
 between two probability models \cite{Bernardo2003}.  When averaged
 over the reference posterior, the reference posterior intrinsic loss
 has a scale which can be mapped onto the displacement with respect to
 the Gaussian mean.  This way, a threshold which corresponds to the
 ``five-sigma'' rule, or any other minimal significance, can be
 defined.  For more details, see \cite{Bernardo2011} and the papers
 cited therein.

 In all searches for new phenomena there is quite a lot of information
 about the background, coming from several auxiliary measurements plus
 Monte Carlo simulations of the known physical processes.  If this
 were not true, nobody would trust a discovery of a new signal based
 on the outcome of the experiment.  Hence we assume that an
 informative prior for $b$ is available which encodes the
 experimenter's degree of belief about the ``reasonable'' range of $b$
 and its ``most likely'' values, written in the form of a Gamma
 density (the conjugate prior of the Poisson model):
\begin{equation}\label{eq-gamma}
 p(b) = \text{Ga}(b|\alpha,\beta)
      = \frac{\beta^\alpha}{\Gamma(\alpha)}
        \, b^{\alpha-1} \, e^{-\beta b}
\end{equation}
 with shape parameter $\alpha>0$ and rate parameter $\beta>0$ (or
 scale parameter $\theta=1/\beta>0$).  In the simple (but frequent)
 case in which there is limited prior knowledge about $b$ and only its
 expectation and variance are known, the Gamma parameters are
 determined by imposing $E[b] = \alpha/\beta$ and $V[b] =
 \alpha/\beta^2$.

 Incidentally, one can note that it is quite common to use a Gaussian
 distribution to model the uncertainty on the yield of the background
 processes, although strictly speaking this is not the correct
 solution.  The reason is that the density function should be defined
 on the domain of the parameter, which in this case is
 $b\in\mathbb{R}^{+}$, whereas the Gaussian distribution is defined
 over the whole real axis.  Hence, the normal distribution needs to be
 truncated at zero, which implies renormalizing it and recomputing its
 mean and standard deviation (when $\text{N}(x;\mu,\sigma)$ is
 truncated at zero, $\mu$ remains the peak position but is no more
 equal to the mean, and $\sigma$ is related to the right-width but is
 no more the standard deviation).  When the distance (in units of
 $\sigma$) between the peak and the origin is big enough, in practice
 one can proceed as if no truncation were necessary.  However, this is
 often not the case, such that truncation does need to be considered.
 Because the domain of a random variable is the very first ingredient
 in the specification of a probability model, it appears more
 reasonable to limit the search for the background prior to the
 functions which are defined on the positive real semiaxis only.  Two
 reasonable choices are the log-normal distribution and the Gamma
 density.  The first may have a theoretical motivation when the
 uncertainty is attributed to the fluctuations of several additional
 independent contributions whose \emph{scale} is uncertain.  However,
 a Gamma density can mimic a log-normal distribution very well (and
 also a normal distribution when the latter is a good approximation)
 and has the advantage of belonging to the family of conjugate priors
 for the Poisson model, which implies that the posterior also belongs
 to the same family.  This makes the choice of a conjugate prior most
 convenient, because simple relations exist between the parameters of
 the prior and posterior Gamma densities, which only depend on the
 observed data.  It is worth emphasizing that this choice does not set
 any limit to the shape of the prior: a linear combination of Gamma
 densities can reproduce any shape.  In this case, the posterior will
 be a linear combination, with the same weights, of the solutions
 which correspond to any individual Gamma density appearing in the
 prior, thanks to the linearity of the Bayes' theorem
 (\ref{eq-bayes-theorem}).

 Often, in particle-physics experiments the selection which is
 performed on the events has the goal of suppressing as much as
 possible the background contribution, while keeping a high efficiency
 for the signal.  This often means that the background estimate has a
 sizable uncertainty, because the background events surviving the
 selection fall in the tails of the distributions of the physical
 observables, which would require the generation of enormous numbers
 of events to be well reproduced by Monte Carlo simulations.  Hence
 the marginal prior density is often quite broad.  In the opposite
 situation in which $b$ is perfectly known, the prior~(\ref{eq-gamma})
 degenerates in a $\delta$ function, which happens when
 $\alpha,\beta\to\infty$ while keeping $E[b]=\alpha/\beta=b_{0}$
 constant (it is sufficient to set $\beta=\alpha/b_{0}$, which implies
 $V[b]=b_{0}/\alpha \to 0$).  In this case, it is easy to show
 (appendix~\ref{sec-delta}) that the reference prior for $s$ is
 Jeffreys' prior for $s'=s+b_{0}$ (i.e.~$b_{0}$ simply redefines the
 origin of the random variable): $\pi(s) \propto (s+b_{0})^{-1/2}$.
 In the following, we address the general problem with any value of
 $\alpha,\beta>0$.

 Because no prior knowledge is assumed for the signal parameter $s$, a
 reference prior $\pi(s)$ for $s$ is desired, such that the joint
 prior can be written in the form $p(s,b) = p(s)\,p(b|s) =
 \pi(s)\,\text{Ga}(b|\alpha,\beta)$.  Two techniques for finding the
 reference prior when the conditional density of the other parameter
 is known are explained by Sun and Berger \cite{Sun1998}, although
 only the second one (``Option 2'' in their paper) is applied here.
 Such technique can be summarized as follows.  The starting point is
 the marginal model $p(k|s)$, specifying the probability of counting
 $k\ge0$ events in the hypothesis that the signal yield is $s\ge0$
 with the assumed knowledge about the background contribution:
\begin{equation}\label{eq-marginal-model}
  p(k|s) = \int_0^\infty \text{Poi}(k|s+b) \,
                         \text{Ga}(b|\alpha,\beta) \di b
 \;.
\end{equation}
 Next, this model is used to compute the Fisher's information
\begin{equation}\label{eq-fisher-info}
\begin{split}
  I(s) &= E\left[ \left( \frac{\partial}{\partial s}
            \log p(k|s) \right)^{\!2\,} \right] 
\\
       &= - E\left[ \left( \frac{\partial^2}{\partial s^2}
            \log p(k|s) \right) \right] 
\end{split}
\end{equation}
 where the last expression is valid for asymptotically normal models,
 as in our case.  Finally, the reference prior for $s$ is then $\pi(s)
 \propto |I(s)|^{1/2}$ \cite{Sun1998}.

 In the rest of the paper, these steps are considered in sequence.
 Section~\ref{sec-marginal-model} illustrates the properties of the
 marginal model $p(k|s)$, which is used in
 section~\ref{sec-fisher-info} to compute the Fisher's information
 $I(s)$.  Finally, the reference prior $\pi(s)$ is computed in
 section~\ref{sec-ref-prior} and few examples of its application are
 shown in section~\ref{sec-posterior}.


\section{The marginal model}\label{sec-marginal-model}

 Here we find an explicit formula for the marginal model
 (\ref{eq-marginal-model}), in terms of the Poisson-Gamma mixture
 \cite{BayesianTheory1994,Bernardo2009b}
\begin{equation}\label{eq-poisson-gamma}
\begin{split}
  P(k|\alpha,\beta) 
               &= \int_0^\infty \text{Poi}(k|\theta) \,
                               \text{Ga}(\theta| \alpha, \beta)
                               \di\theta
\\
               &= \frac{\beta^{\alpha} \, \Gamma(\alpha+k)}
	              {k!\,\Gamma(\alpha) \, (1+\beta)^{\alpha+k}}
\end{split}
\end{equation}
 a result which can be obtained by means of the Gamma integral
\begin{equation*}
  \Gamma(a) = \int_0^\infty x^{a-1} e^{-x} \di x  \; .
\end{equation*}

 First, we note that
\begin{equation}
\begin{split}
  \text{Poi}(k|s+b) &= e^{-s-b} \, \frac{(s+b)^k}{k!}
\\
      &=  \frac{e^{-s-b}}{k!} \sum_{n=0}^{k} \binom{k}{n} \, s^{k-n} \, b^{n}
\\
      &=  e^{-s-b} \sum_{n=0}^{k} \frac{s^{k-n} \, b^{n}}{n!\,(k-n)!}
\end{split}
\end{equation}
 which allows to rewrite (\ref{eq-marginal-model}) in the form
\begin{equation}
\begin{split}
  p(k|s) &= \sum_{n=0}^{k} \frac{e^{-s} \, s^{k-n}}{(k-n)!} 
           \int_0^\infty \frac{b^n\,e^{-b}}{n!}
              \, \text{Ga}(b| \alpha, \beta) \di b
  \\
         &= \sum_{n=0}^{k} \frac{e^{-s} \, s^{k-n}}{(k-n)!} 
           \int_0^\infty \text{Poi}(n|b)
              \, \text{Ga}(b| \alpha, \beta) \di b
\end{split}
\end{equation}
 which, with the help of (\ref{eq-poisson-gamma}), becomes
\begin{equation}\label{eq-marg-mod}
\begin{split}
  p(k|s) &= \left( \frac{\beta}{1+\beta} \right)^{\!\alpha}
            \sum_{n=0}^{k} 
            \frac{\Gamma(n+\alpha) / \Gamma(\alpha)}
		 {n! \, (1+\beta)^{n}}
            \,   \frac{e^{-s} \,s^{k-n}}{(k-n)!}
  \\
         &= \left( \frac{\beta}{1+\beta} \right)^{\!\alpha}
            \sum_{n=0}^{k} \frac{(\alpha)^{(n)}}{n! \, (1+\beta)^{n}}
	    \, \text{Poi}(k-n|s)
  \; .
\end{split}
\end{equation}

 The ``rising factorial'' (or Pochammer function)
\(
   (\alpha)^{(n)}
     = \Gamma(n+\alpha) / \Gamma(\alpha)
     = \alpha (\alpha+1) (\alpha+2) \cdots (\alpha+n-1)
\) is defined as
\[
 (\alpha)^{(n)} = 
\begin{cases}
  1                                &\text{if}\;\, n=0 \\
  \alpha                           &\text{if}\;\, n=1 \\
  (\alpha)^{(k-1)} (\alpha+k-1)    &\text{if}\;\, n=k>1 \\
\end{cases}
\]
 and the ratio $(\alpha)^{(n)}/n!$ defines the binomial coefficient
 $\binom{\alpha+n-1}{n}$ with a upper real parameter, such that the
 the marginal model (\ref{eq-marg-mod}) can be also written in the
 final form
\begin{equation}\label{eq-marg-mod-2}
  p(k|s) = \left( \frac{\beta}{1+\beta} \right)^{\!\alpha}
           e^{-s} \, f(s;k,\alpha,\beta)
  \; .
\end{equation}
 where the polynomial
\begin{equation}\label{eq-f}
 f(s;k,\alpha,\beta) = \sum_{n=0}^{k} \binom{\alpha+n-1}{n}
                       \frac{s^{k-n}}{(k-n)! \, (1+\beta)^{n}}
\end{equation}
 has explicit forms given in table~\ref{tab-marg-mod}. 

 The function $f(s;k,\alpha,\beta)$ has some interesting properties
 which are useful in the following treatment.  The proofs are given in
 appendix~\ref{sec-proofs}.  Appendix~\ref{sec-code} provides
 additional information which is useful when writing code to evaluate
 this polynomial.

\begin{prop}\label{prop1}
 For each $\alpha,\beta>0$ and $s\ge0$ the series
\(
  \sum_{k=0}^{\infty} f(s;k,\alpha,\beta)
\)
 converges to
\[
  \sum_{k=0}^{\infty} f(s;k,\alpha,\beta) =
   \left( \frac{1+\beta}{\beta} \right)^{\!\alpha} e^{s} \; .
\]
\end{prop}
 This ensures that the the marginal model (\ref{eq-marg-mod-2}) is
 properly normalized.

\begin{prop}\label{prop2}
 The $n$-th partial derivative of $f(s;k,\alpha,\beta)$ with respect
 to $s$ is
\begin{equation}\label{eq-f-derivatives}
  f^{(n)}(s;k,\alpha,\beta)
   =
\begin{cases}
  0                      &\text{if}\quad k<n \\
  1                      &\text{if}\quad k=n \\
  f(s;k-n,\alpha,\beta)  &\text{if}\quad k\ge n \\
\end{cases}
\end{equation}
\end{prop}
 Property 2 makes it easy to compute all derivatives, the evaluation
 of a single function being sufficient.

\begin{prop}\label{prop3}
 For each finite $n\ge1$, the sum
\(
  \sum_{k=0}^{\infty} f^{(n)}(s;k,\alpha,\beta)
\)
 also converges to
\(
  e^{s} (1+\beta)^{\alpha} / \beta^{\alpha}
\).
\end{prop}
 This ensures that the the expectation $E[f^{(n)}/f] = 1$ for each $n$.

\begin{figure}[t]
  \centering
  \includegraphics[width=0.6\textwidth]{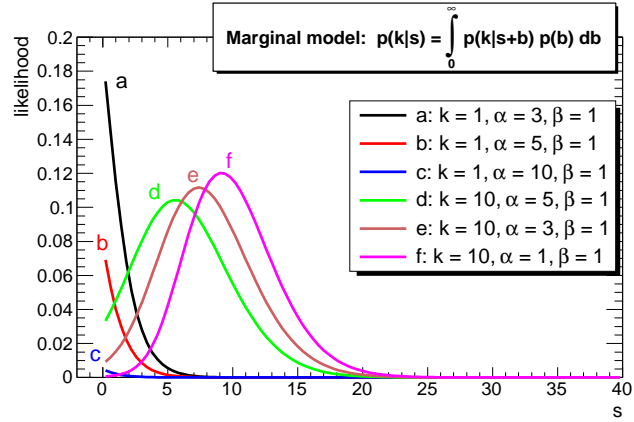}
  \caption{Marginal likelihood for different background priors, with 1
  and 10 observed counts.}
  \label{fig-marg-likel}
\end{figure}

 The marginal likelihood is shown in figure~\ref{fig-marg-likel}, for
 different background parameters, in the case of 1 and 10 observed
 counts.  As shown in details in the appendix~\ref{sec-other-papers},
 the marginal model (\ref{eq-marg-mod-2}) coincides with the
 appropriate limit of the marginal model of the DJP paper and with the
 marginal model of the PPSS paper.

\begin{table}[b]
\begin{tabular}{cl}
 $k$ & \raisebox{0pt}[2ex][1ex]{$f(s;k,\alpha,\beta)$} \\
\hline
  0 \raisebox{0pt}[4ex][2ex]{~} &  1 \\
  1 \raisebox{0pt}[4ex][2ex]{~} & 
\(
\displaystyle
  s + \frac{\alpha}{1+\beta}
\) \\
  2 \raisebox{0pt}[4ex][2ex]{~} & 
\(
\displaystyle
  \frac{s^2}{2} + \frac{s\,\alpha}{1+\beta}
       + \frac{\alpha(\alpha+1)}{2(1+\beta)^2}
\) \\
  3 \raisebox{0pt}[4ex][2ex]{~} & 
\(
\displaystyle
  \frac{s^3}{6} + \frac{s^2 \alpha}{2(1+\beta)}
       + \frac{s\,\alpha(\alpha+1)}{2(1+\beta)^2}
       + \frac{\alpha(\alpha+1)(\alpha+2)}{6(1+\beta)^3}
\) \\
$n>4$ \raisebox{0pt}[4ex][2ex]{~} & 
\(
\displaystyle
  \frac{s^n}{n!} 
       + \frac{s^{n-1}}{(n-1)!} \frac{\alpha}{(1+\beta)}
       + \frac{s^{n-2}}{(n-2)!} \frac{\alpha(\alpha+1)}{2(1+\beta)^2}
       + \frac{s^{n-3}}{(n-3)!} \frac{\alpha(\alpha+1)(\alpha+2)}{3! \, (1+\beta)^3}
       + \cdots
\) \\
 & 
\(
\displaystyle
       \cdots
       + \frac{\alpha(\alpha+1)\cdots(\alpha+n-1)}{n! \, (1+\beta)^n}
\) \\
\end{tabular}

\caption{Explicit forms of $f(s;k,\alpha,\beta)$
 for small values of $k$.}\label{tab-marg-mod} 
\end{table}


\section{The Fisher's information}\label{sec-fisher-info}

 The logarithm of the marginal model (\ref{eq-marg-mod-2}) is
\begin{equation}\label{eq-log-marg-mod}
  \log p(k|s) = \alpha \log\frac{\beta}{1+\beta} - s
              + \log f(s;k,\alpha,\beta)
\end{equation}
 and its derivative with respect to $s$ is
\begin{equation}\label{eq-log-marg-mod-derivative}
  \frac{\partial \log p(k|s)}{\partial s} = -1 
        + \frac{f^{(1)}(s;k,\alpha,\beta)}{f(s;k,\alpha,\beta)}
\end{equation}
 while the second derivative is
\begin{equation}\label{eq-log-marg-mod-2nd-deriv}
  \frac{\partial^2 \log p(k|s)}{\partial s^2} = 
        \frac{f^{(2)}(s;k,\alpha,\beta)}{f(s;k,\alpha,\beta)}
        - \left[ \frac{f^{(1)}(s;k,\alpha,\beta)}{f(s;k,\alpha,\beta)}
	  \right]^2
\end{equation}

 The expectation in the definition of the Fisher's information
 (\ref{eq-fisher-info}) splits in two terms:
\begin{equation}\label{eq-info-expec}
  I(s) = E\left[\frac{[f^{(1)}(s;k,\alpha,\beta)]^2}
                     {[f(s;k,\alpha,\beta)]^2}\right]
       - E\left[\frac{f^{(2)}(s;k,\alpha,\beta)}{f(s;k,\alpha,\beta)}\right]
\end{equation}
 which can be evaluated independently.  The expectation $E[f^{(2)}/f]$ is
 one by virtue of Property~3.

 The other term in (\ref{eq-info-expec}) is
\begin{equation*}
\begin{split}
  E\left[ (f^{(1)}/f)^{2} \right]
        &= \left( \frac{\beta}{1+\beta} \right)^{\!\alpha} e^{-s} \,
           \sum_{k=0}^{\infty} 
           \frac{[f^{(1)}(s;k,\alpha,\beta)]^2}{f(s;k,\alpha,\beta)}
  \\
    \qquad \{ k=0 \Rightarrow & f^{(1)}=0 \;
     \text{hence start from $k=1$; use Property 2}\}
  \\
        &= \left( \frac{\beta}{1+\beta} \right)^{\!\alpha} e^{-s} \,
           \sum_{k=1}^{\infty}
           \frac{[f(s;k-1,\alpha,\beta)]^2}{f(s;k,\alpha,\beta)}
  \\
    \qquad\{ \text{now set} &\; n = k-1 \}
  \\
        &= \left( \frac{\beta}{1+\beta} \right)^{\!\alpha} e^{-s} \,
           \sum_{n=0}^{\infty}
           \frac{[f(s;n,\alpha,\beta)]^2}{f(s;n+1,\alpha,\beta)}
\end{split}
\end{equation*}
 which is easy to implement in a numerical routine, because one only
 needs a single evaluation of $f(s;k,\alpha,\beta)$ for each term in
 the sum (appendix~\ref{sec-code}), in a loop which terminates when
 the current addendum gives a negligible contribution (for example,
 when the ratio between the addendum and the partial sum becomes less
 than $10^{-6}$, as it was done in the examples considered in the next
 section).

 Finally, the Fisher's information is
\begin{equation}\label{eq-fisher-info-2}
  I(s) = \left( \frac{\beta}{1+\beta} \right)^{\!\alpha} e^{-s} \,
           \sum_{n=0}^{\infty}
           \frac{[f(s;n,\alpha,\beta)]^2}{f(s;n+1,\alpha,\beta)}
       - 1
\end{equation}
 the same result being obtained when computing the expectation of the
 square of the first derivative (\ref{eq-log-marg-mod-derivative}) of
 the marginal log-likelihood, because $E[f^{(1)}/f]=1$ by virtue of
 Property~3.  Appendix~\ref{sec-delta} shows that $I(s)$ gives the
 correct reference prior (Jeffreys' prior) when there is certain prior
 knowledge about the background yield.


\section{The reference prior}\label{sec-ref-prior}

 The reference prior for $s$ is $\pi(s) \propto |I(s)|^{1/2}$
 \cite{Sun1998}.  From equation~(\ref{eq-fisher-info-2}) one obtains
\begin{equation}\label{eq-sqrt-fisher-info}
  |I(s)|^{1/2}  =  \left|
       \left( \frac{\beta}{1+\beta} \right)^{\!\alpha} e^{-s} \,
           \sum_{n=0}^{\infty}
           \frac{[f(s;n,\alpha,\beta)]^2}{f(s;n+1,\alpha,\beta)}
       - 1
                 \right|^{1/2}
\end{equation}
 which only requires the evaluation of the function
 $f(s;n,\alpha,\beta)$ (see appendix~\ref{sec-code}) to be computed
 once per cycle.

 When $s\to\infty$ the function~(\ref{eq-sqrt-fisher-info}) is not
 decreasing fast enough to make it integrable over the positive real
 axis.  In the asymptotic regime, the leading term in the polynomial
 $f(s;k,\alpha,\beta)$ is proportional to $s^k$, which means that each
 addendum in the sum diverges as a polynomial of degree $s^{n-1}$, and
 the exponential $e^{-s}$ is not sufficient to ensure that the sum
 goes to zero for $s\to\infty$ (easier to check in log scale:
 $\log(e^{-s}\,s^n) = -s + n\log s \to\infty$ when both $n$ and $s$ go
 to $\infty$).  This means that $\pi(s)$ is not a proper prior: it
 cannot represent somebody's degree of belief.  In the framework of
 the Bayesian reference analysis, the use of improper reference priors
 is admitted, provided that the reference posterior is a proper
 probability density (which is the case here).

\begin{figure*}[t!]
\begin{minipage}[t]{0.5\textwidth}
  \centering
  \includegraphics[width=0.95\textwidth]{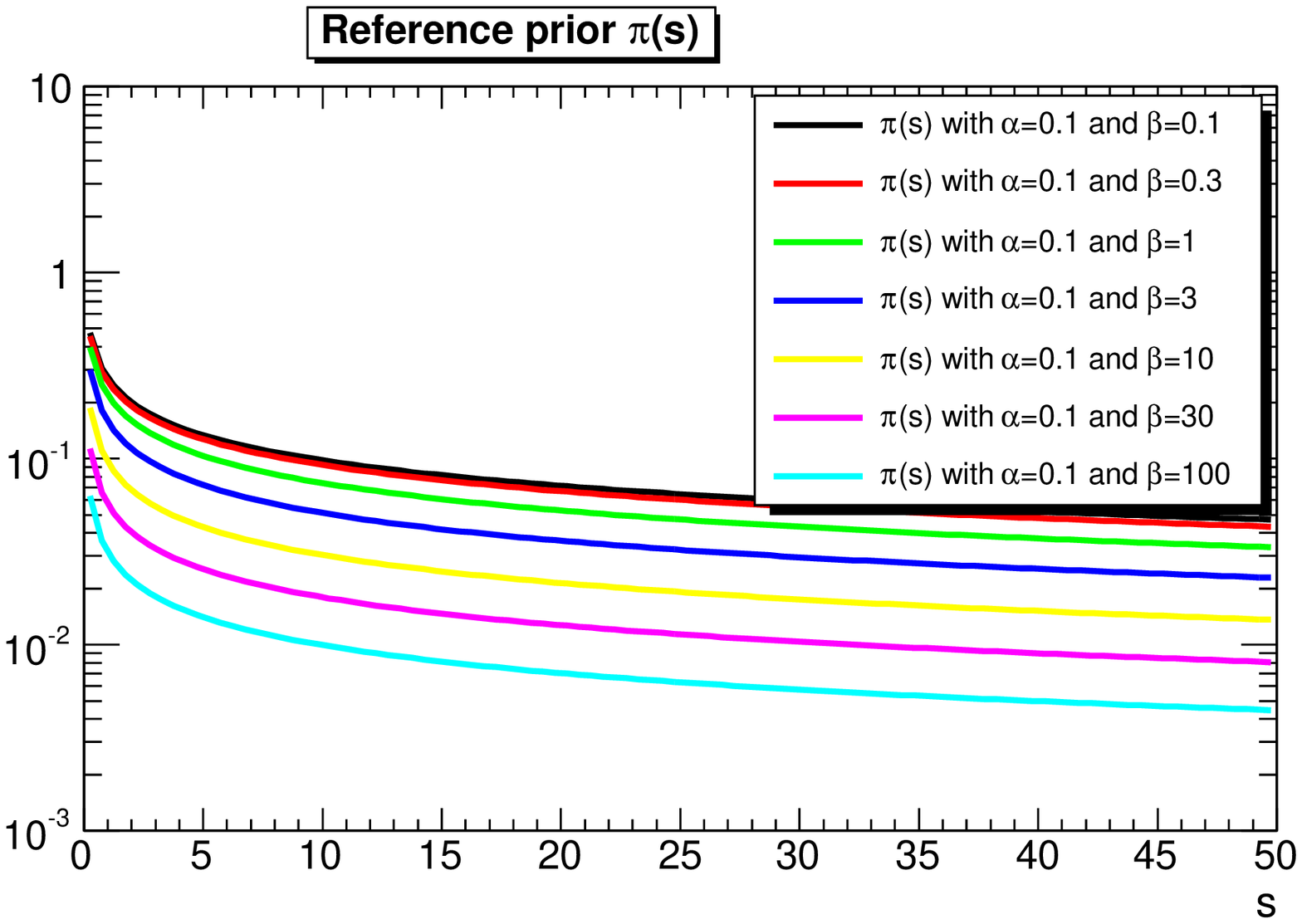}
  \includegraphics[width=0.95\textwidth]{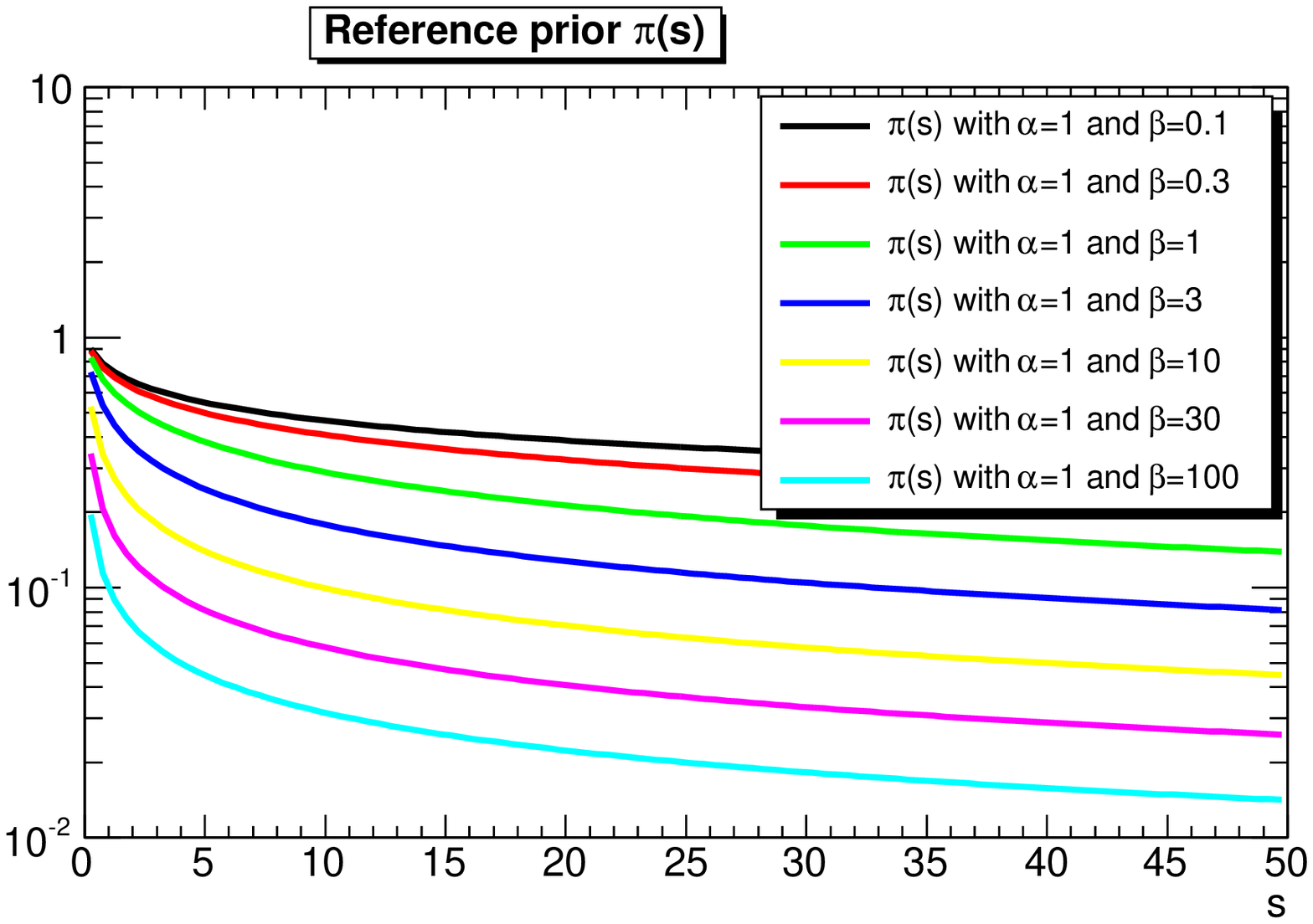}
\end{minipage}%
\begin{minipage}[t]{0.5\textwidth}
  \centering
  \includegraphics[width=0.95\textwidth]{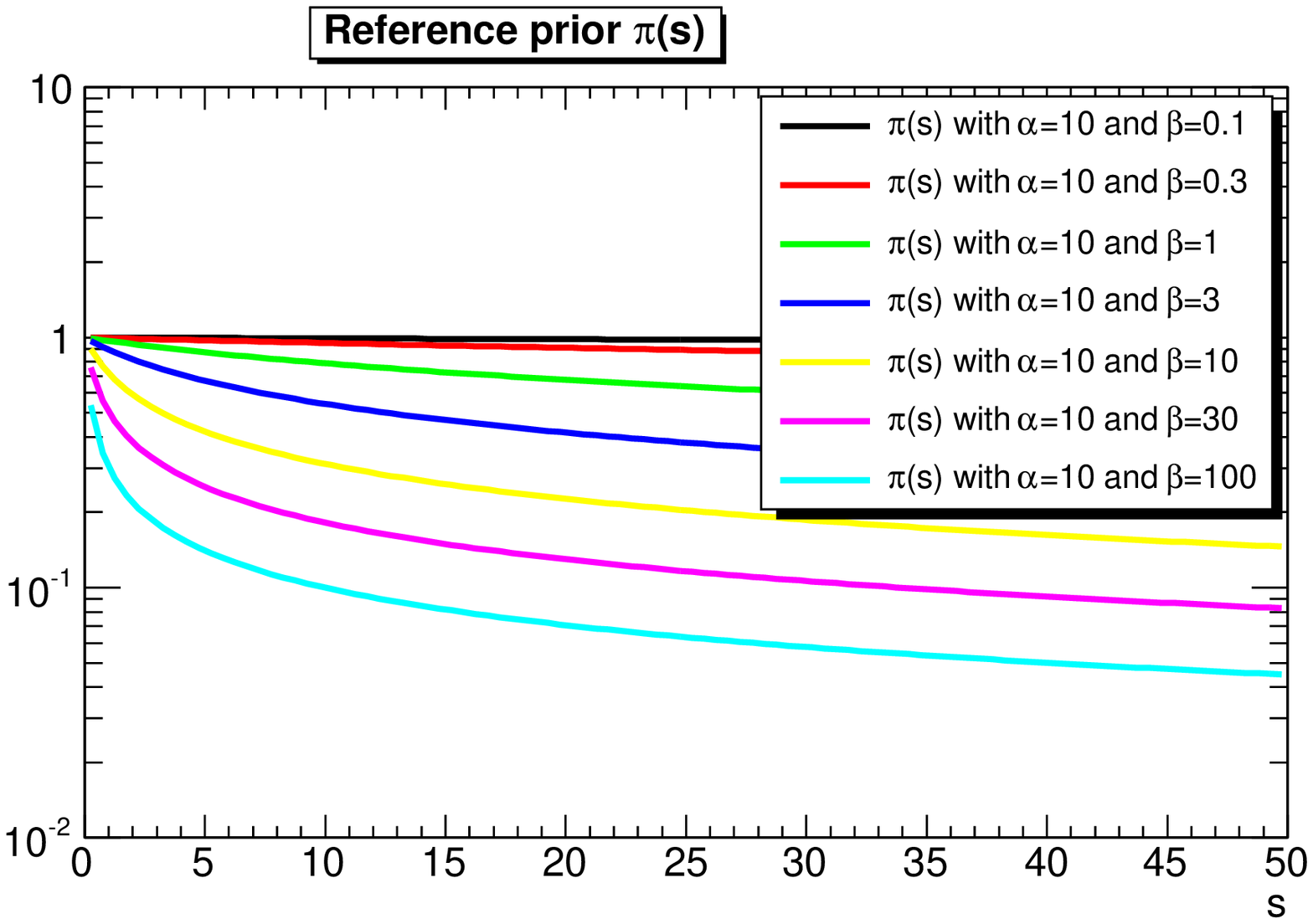}

\vspace{3em}

  \caption{Reference prior for the signal parameter $s$ obtained with
  background parameters $\alpha=0.1$ (top-left), $\alpha=1$
  (bottom-left), $\alpha=10$ (top-right) and $\beta=0.1, 0.3, 1, 3,
  10, 30, 100$.  The top-down sequence of the functions is the same as
  the list in the legend of each plot.}
  \label{fig-refer-prior}
\end{minipage}%
\end{figure*}

 For $s\to0$ the polynomial $f(s;n,\alpha,\beta)$ reduces to a
 constant
\begin{equation}\label{eq-f-zero}
\begin{split}
  f(0;n,\alpha,\beta)
          &= \binom{\alpha+n-1}{n} (1+\beta)^{-n}
\\
          &= \frac{\alpha+n-1}{n(1+\beta)} \, f(0;n-1,\alpha,\beta)
\end{split}
\end{equation}
 which goes to zero for $n\to\infty$ fast enough to make the series in
 equation~(\ref{eq-sqrt-fisher-info}) converge.  The function
 $|I(s)|^{1/2}$ has its single maximum at zero, hence a possible
 definition of the reference prior for the signal is
\begin{equation}\label{eq-ref-prior-signal}
  \pi(s) = \frac{|I(s)|^{1/2} }{|I(0)|^{1/2} }
\end{equation}
 which is a monotonically decreasing function attaining its maximum
 value of one at zero.  The reference prior $\pi(s)$ is shown in
 figure~\ref{fig-refer-prior} for different choices of background
 parameters.  The maximum at $s=0$ becomes more pronounced for small
 values of $\alpha$ and large values of $\beta$, whereas large values
 of $\alpha$ and small values of $\beta$ make the prior flat.  As
 expected, combinations which have similar background expectation and
 variance give almost the same curve.  In figure~\ref{fig-refer-prior}
 such examples are: the $(\alpha,\beta)$ pairs $(0.1,3)$ and $(1,30)$,
 which have $E[b]=0.03$ and $E[b]=0.033$, both with $V[b]=0.011$; the
 pairs $(0.1,10)$ and $(1,100)$, both with $E[b]=0.01$, which have
 $V[b]=0.001$ and $V[b]=0.0001$; the pairs $(1,10)$ and $(10,100)$,
 both with $E[b]=0.1$, which have $V[b]=0.01$ and $V[b]=0.001$.

\section{The marginal reference posterior}\label{sec-posterior}

 The joint prior for our model is the improper function $\pi(s,b) =
 \pi(s) \, p(b)$ where $\pi(s)$ is defined by
 equation~(\ref{eq-ref-prior-signal}) (or any other function which is
 proportional to $|I(s)|^{1/2}$) and $p(b)$ is the Gamma
 density~(\ref{eq-gamma}) which encodes the experimenter's prior
 degree of belief about the background.

 The joint reference posterior is proportional to
\(
  p(s,b|k) \propto \text{Poi}(k|s+b) \, p(b) \, \pi(s)
\)
 and the joint marginal posterior for $s$ is obtained after
 integration over $b$.  Finally, because $\pi(s)$ does not explicitly
 depend on $b$, the marginal posterior is proportional to the product
 of the reference prior (\ref{eq-ref-prior-signal}) and the marginal
 likelihood (\ref{eq-marg-mod-2}):
\begin{equation}\label{eq-marg-post}
  p(s|k) \propto \left( \frac{\beta}{1+\beta} \right)^{\!\alpha}
                 e^{-s} \, f(s;k,\alpha,\beta) \, \pi(s) \; .
\end{equation}

\begin{figure*}
\begin{minipage}[b]{0.5\textwidth}
 \centering
 \begin{tabular}{ccrr}
 $E[b]$ & rel.~unc. & $\alpha$ & $\beta$  \\
 \hline
  2     &   0.1     &  100     &  50.0    \\
  2     &   0.2     &   25     &  12.5    \\
  2     &   0.5     &    4     &   2.0    \\
  2     &   1.0     &    1     &   0.5    \\
 \end{tabular}
 \caption{Parameters for the background priors.}
 \label{tab-bkg-priors-Eb2}
\end{minipage}%
\begin{minipage}[b]{0.5\textwidth}
 \centering
 \includegraphics[width=\textwidth]{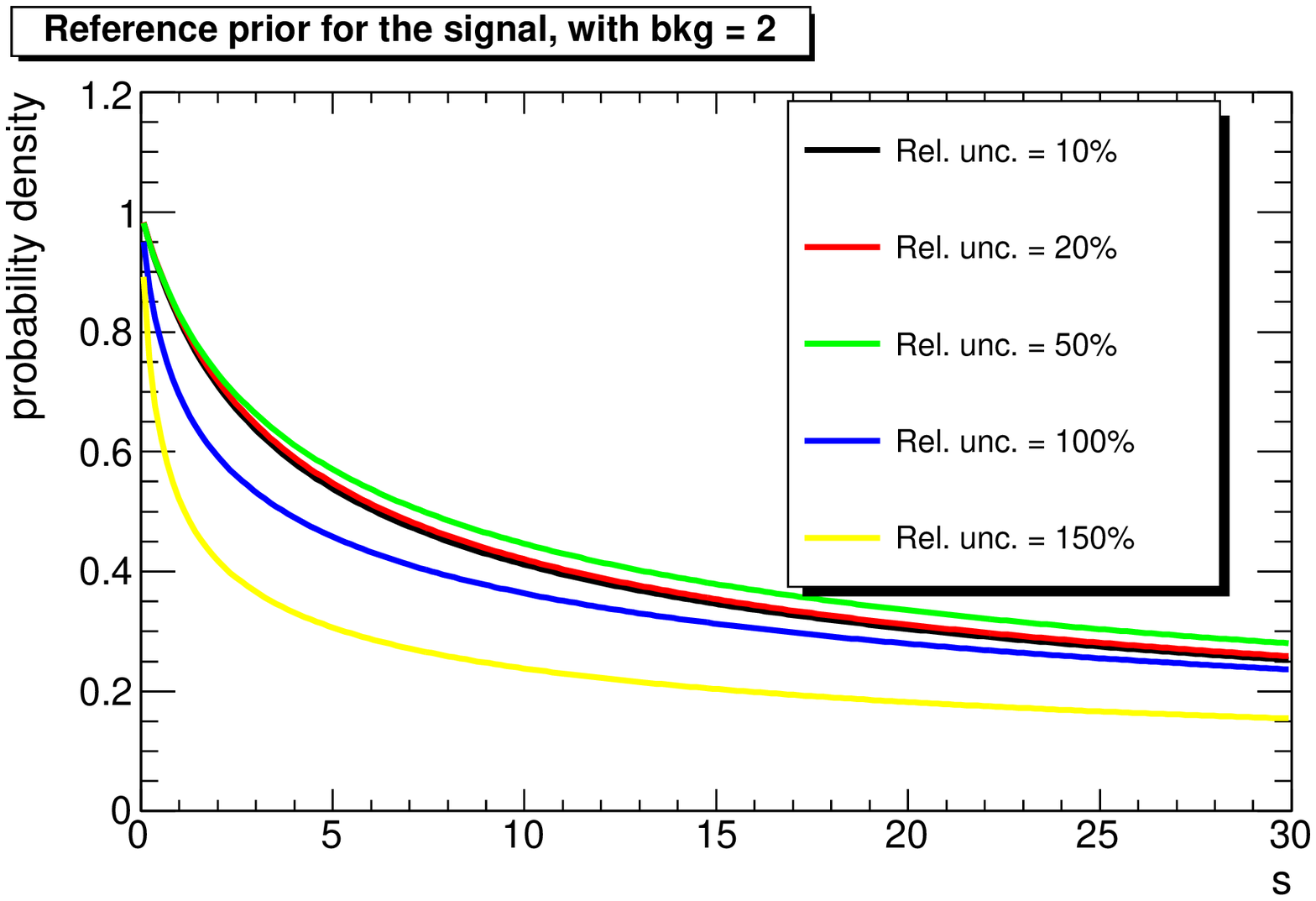}
 \caption{The background prior distributions.}
 \label{fig-bkg-priors-Eb2}
\end{minipage}%
\end{figure*}

\begin{figure*}[t!]
\begin{minipage}[b]{0.5\textwidth}
 \centering
 \includegraphics[width=\textwidth]{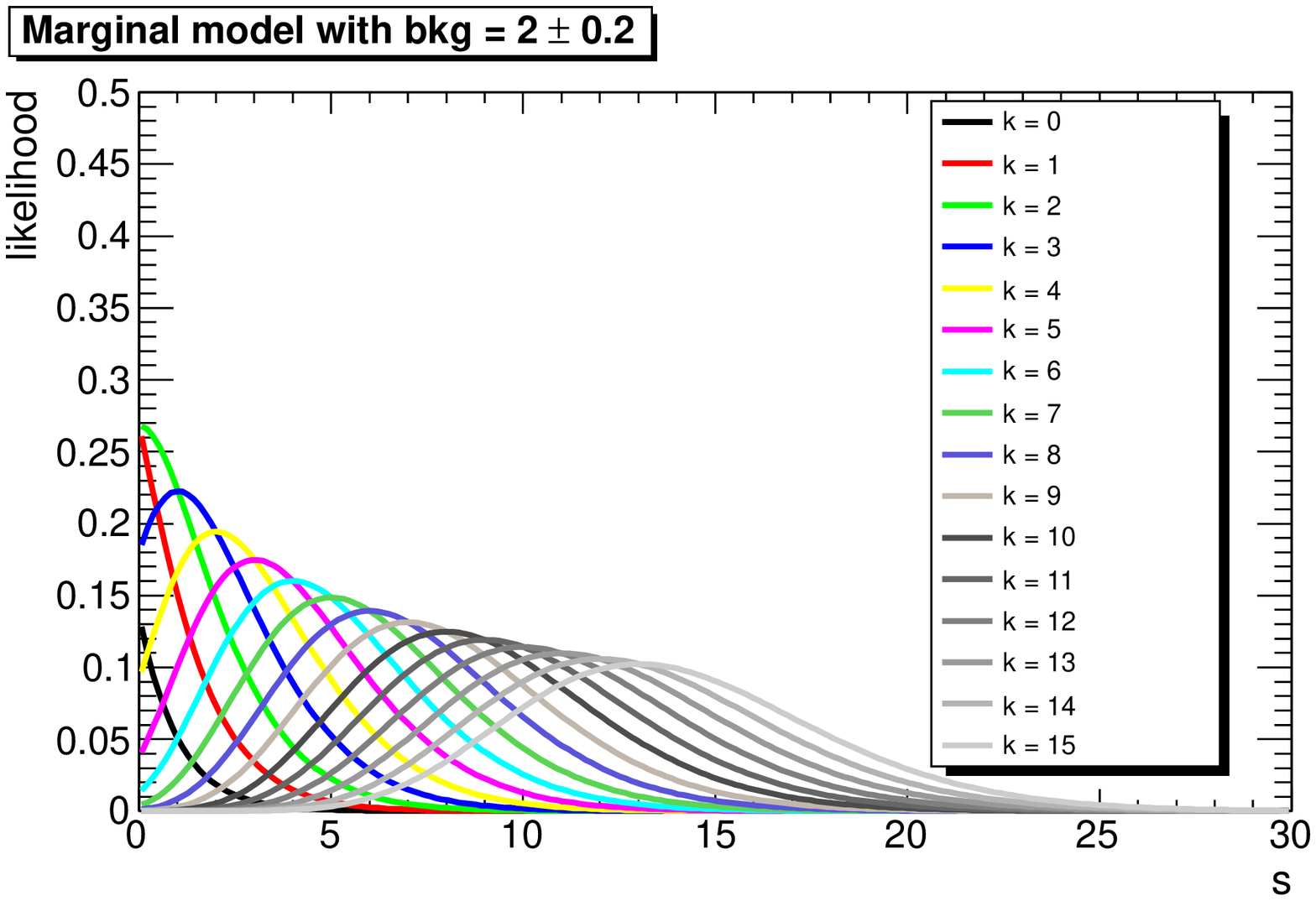}
 \includegraphics[width=\textwidth]{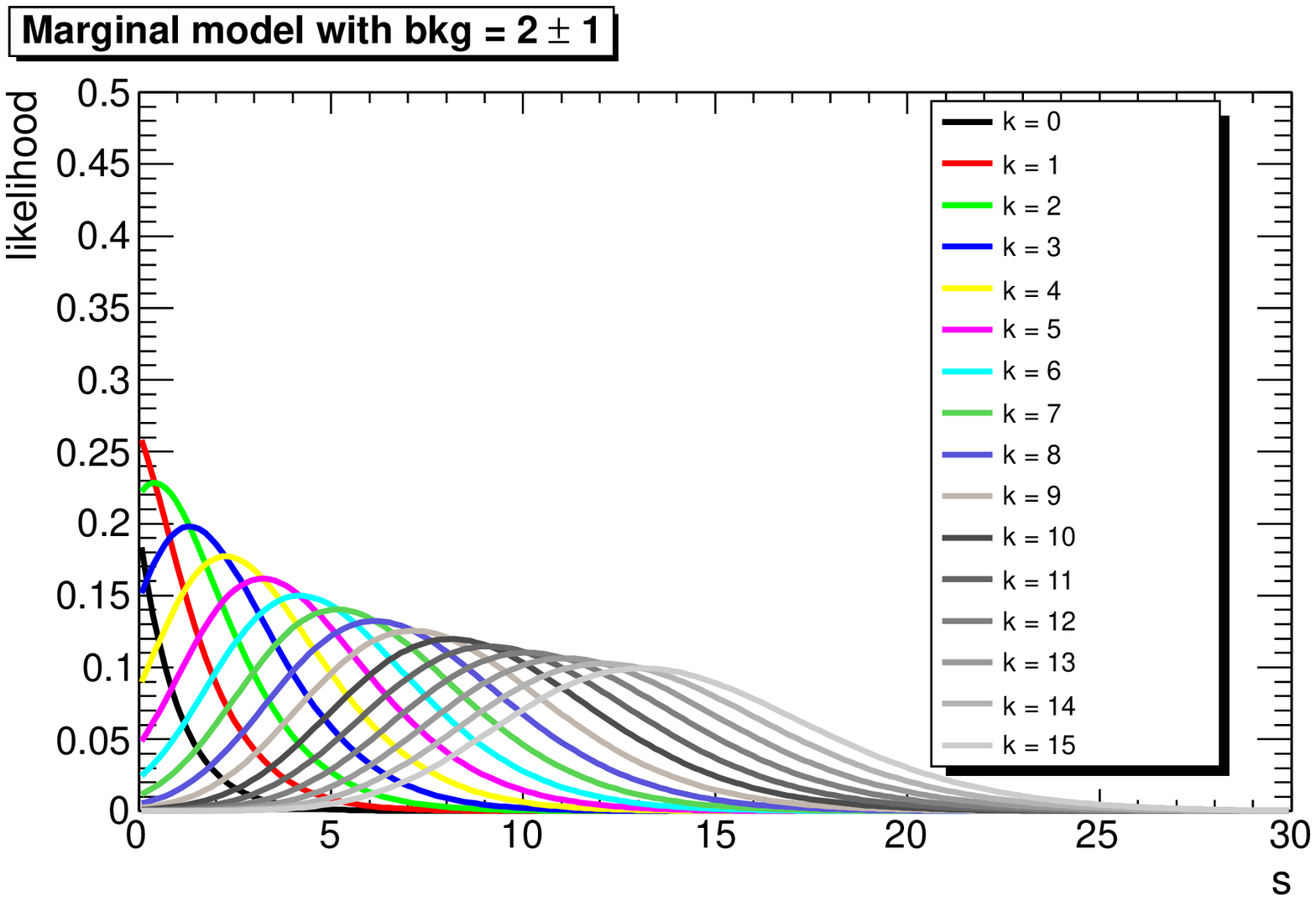}
\end{minipage}%
\begin{minipage}[b]{0.5\textwidth}
 \centering
 \includegraphics[width=\textwidth]{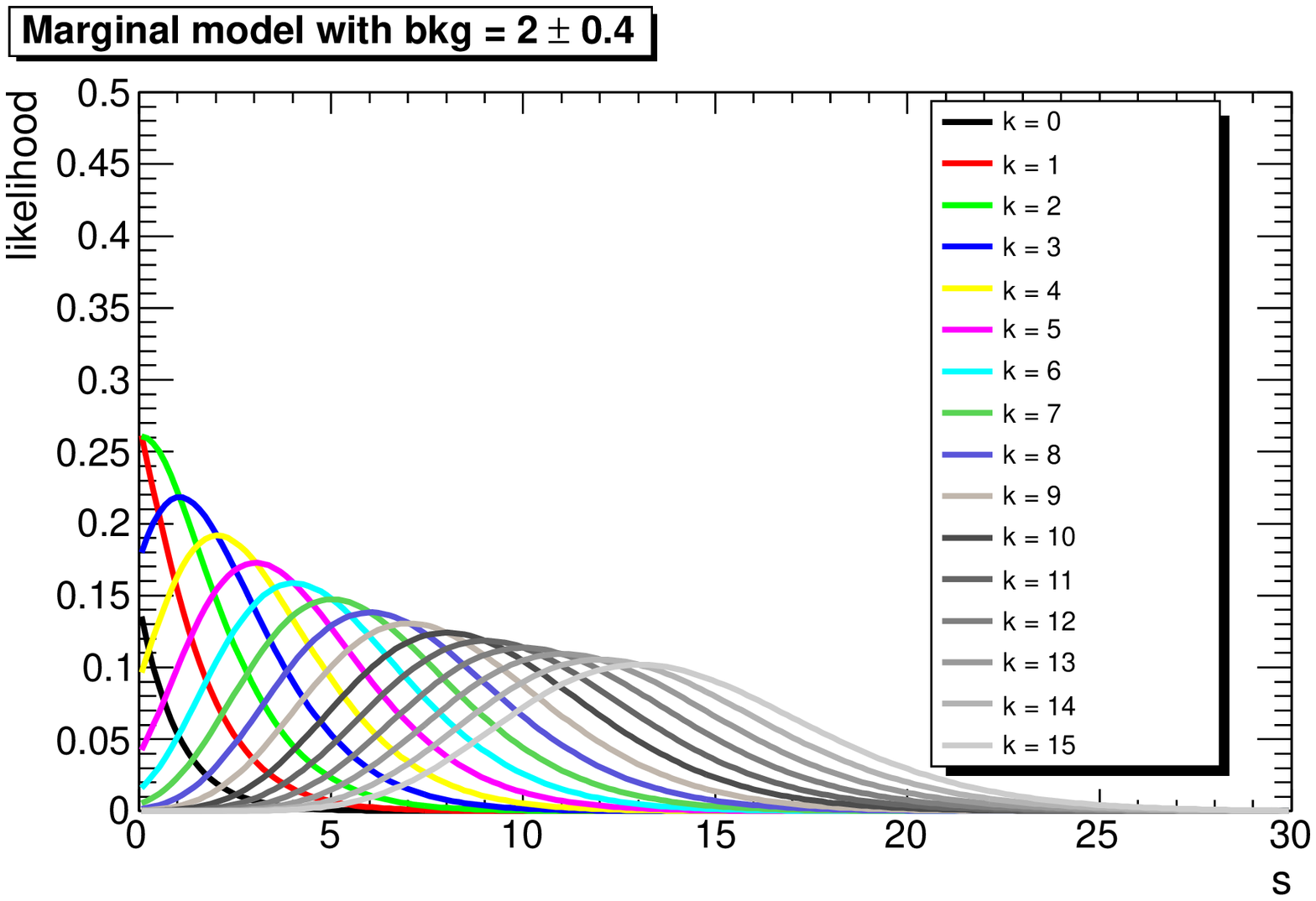}
 \includegraphics[width=\textwidth]{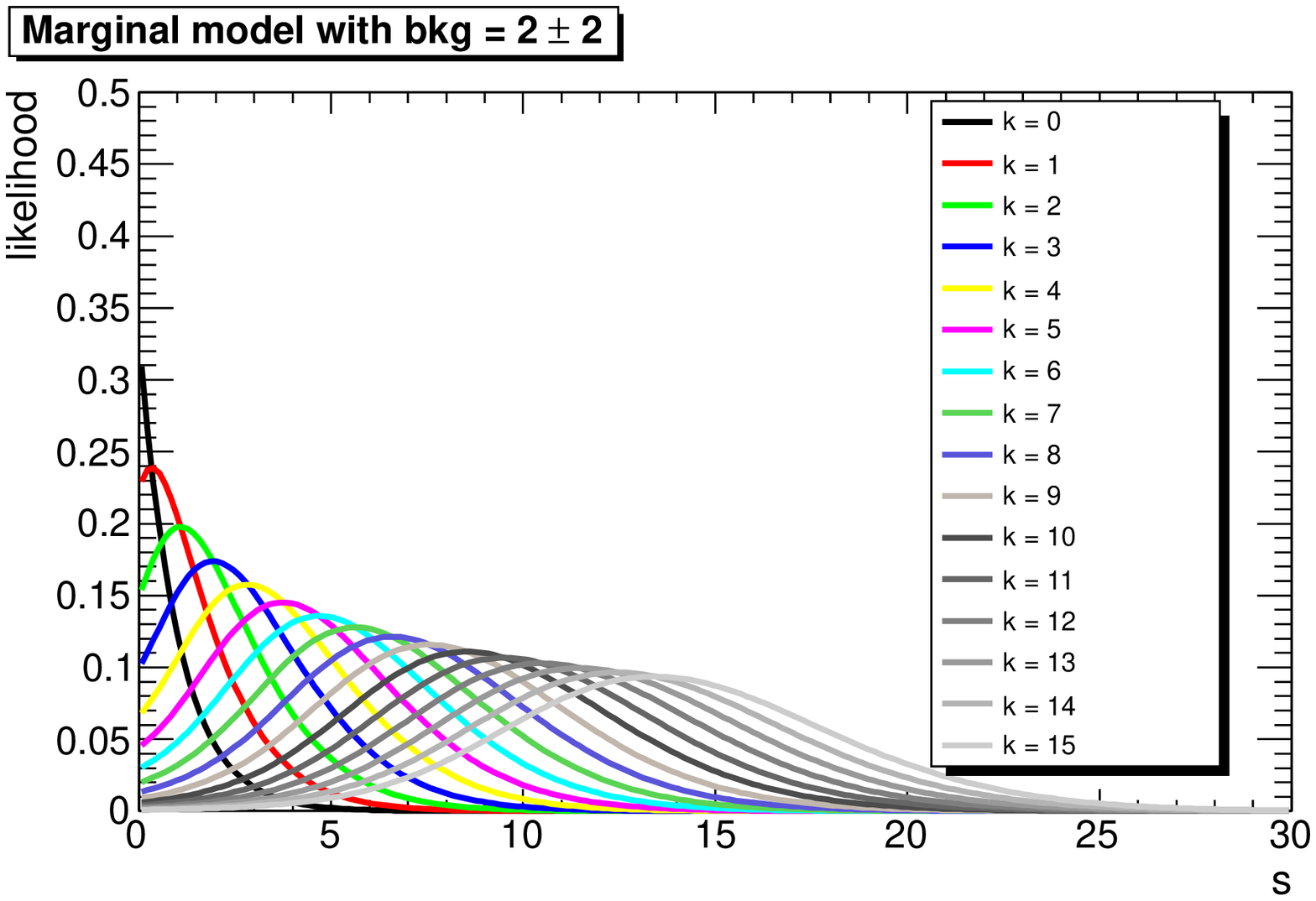}
\end{minipage}%
 \caption{Marginal likelihood for the true signal, for different
   background priors ($E[b]=2$ with relative uncertainty 10\%, 20\%,
   50\%, and 100\% from top-left to bottom-right) and sample sizes
   ($k=0,1,\ldots,15$).}
 \label{fig-marg-likelihoods}
\end{figure*}

\begin{figure*}[t!]
\begin{minipage}[b]{0.5\textwidth}
 \centering
 \includegraphics[width=\textwidth]{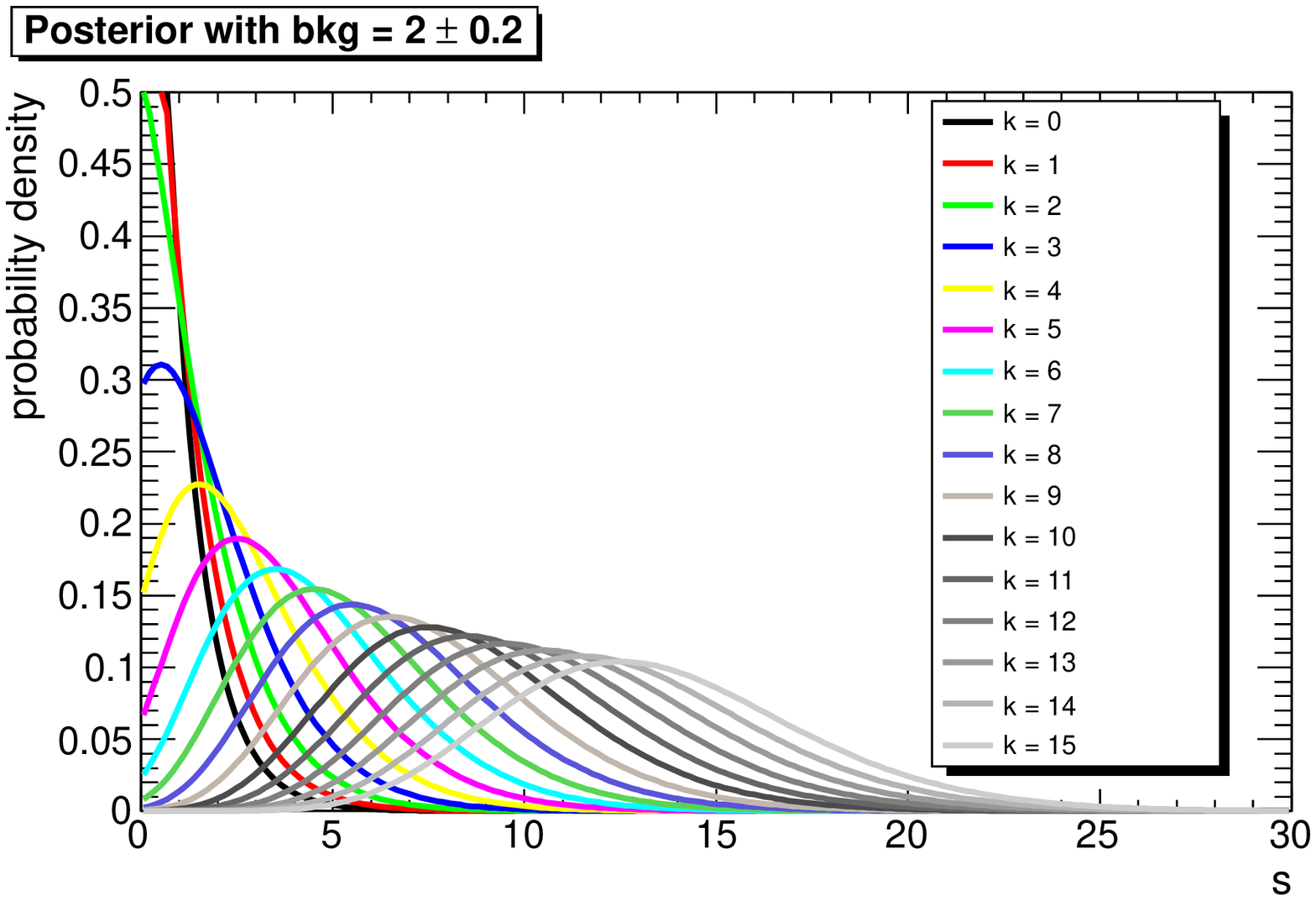}
 \includegraphics[width=\textwidth]{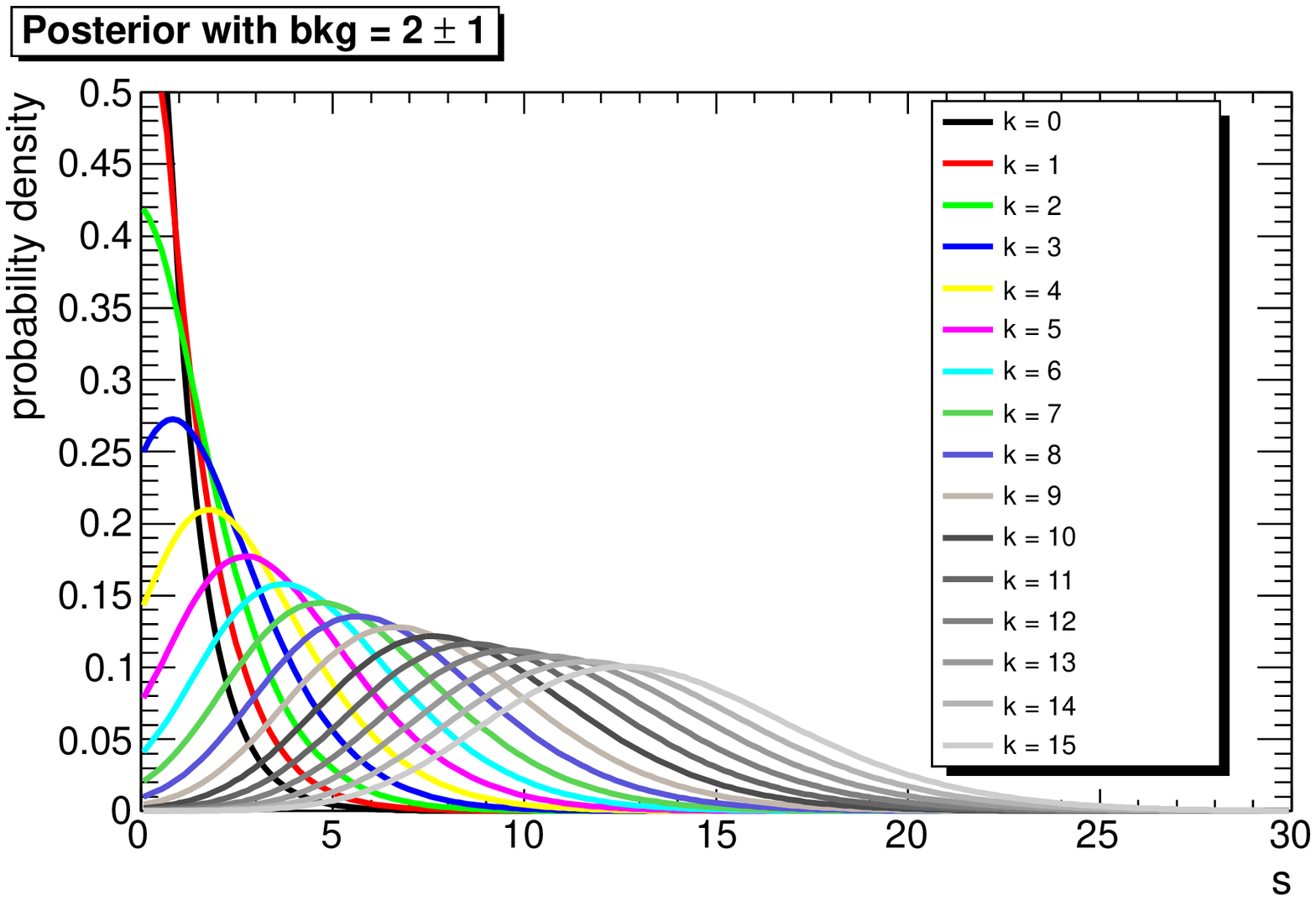}
\end{minipage}%
\begin{minipage}[b]{0.5\textwidth}
 \centering
 \includegraphics[width=\textwidth]{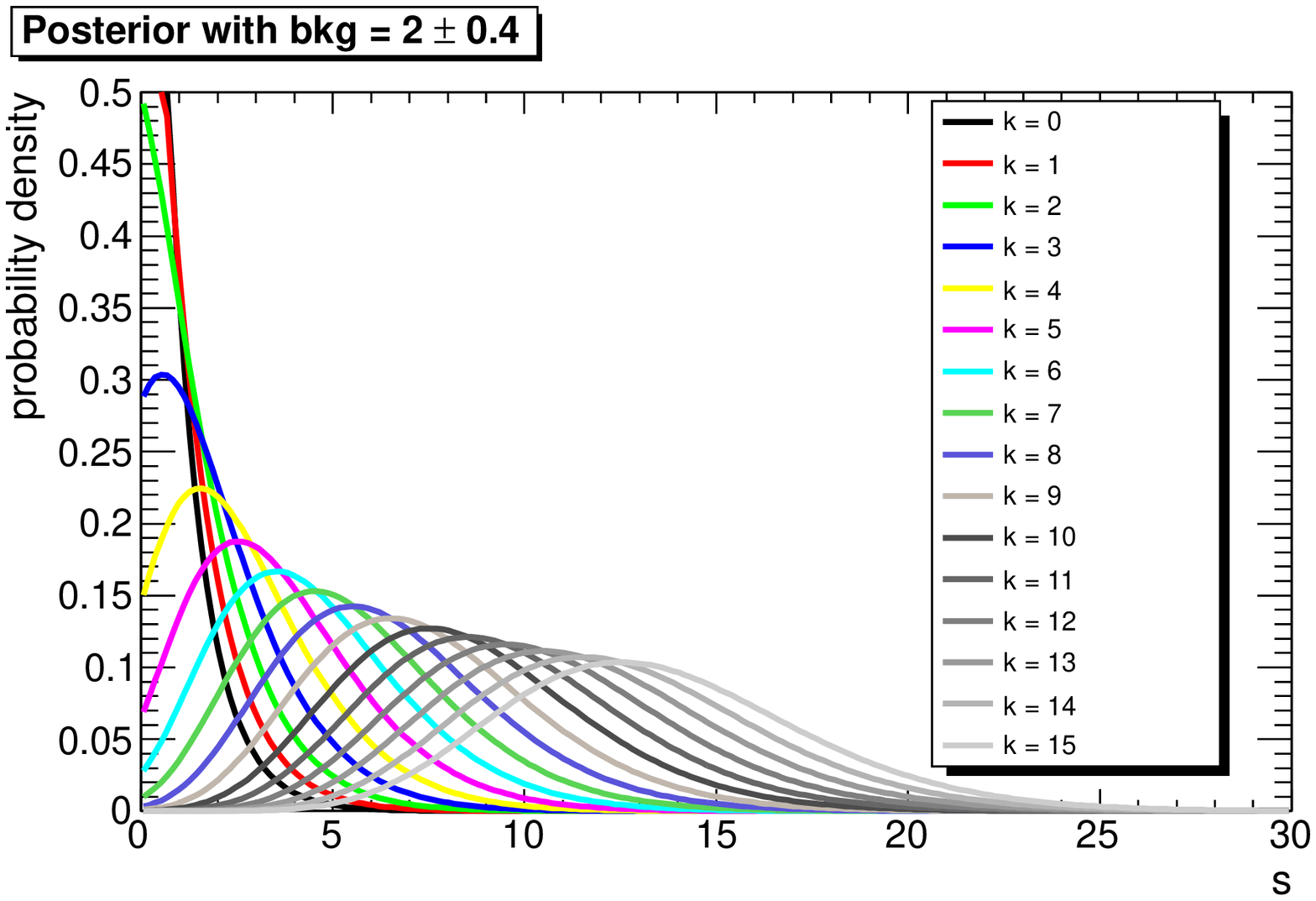}
 \includegraphics[width=\textwidth]{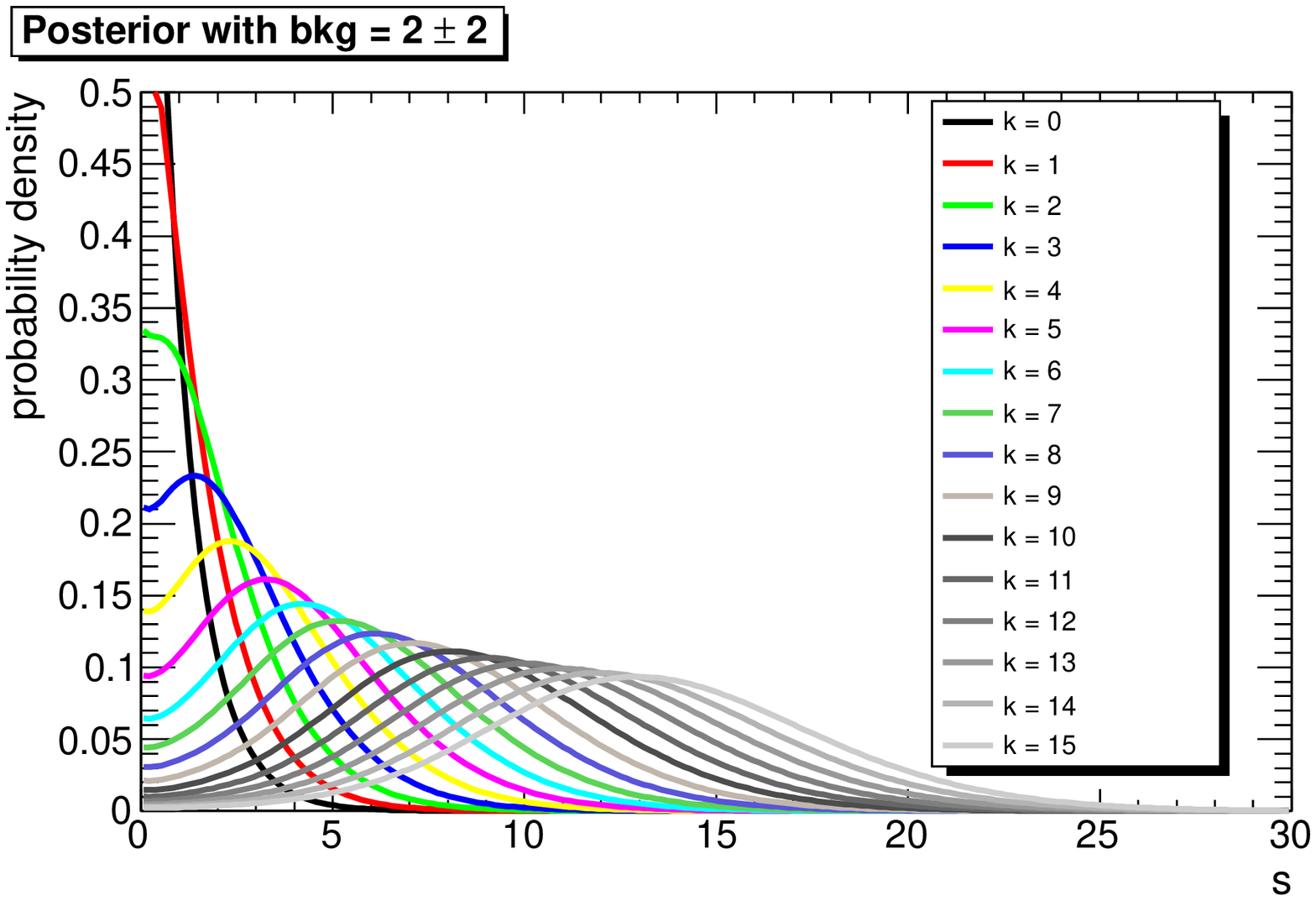}
\end{minipage}%
 \caption{Marginal posteriors for the signal, for different background
   priors ($E[b]=2$ with relative uncertainty 10\%, 20\%, 50\%, and
   100\% from top-left to bottom-right) and sample sizes
   ($k=0,1,\ldots,15$).}
 \label{fig-marg-posteriors}
\end{figure*}

 In order to illustrate the properties of the marginal reference
 posterior (\ref{eq-marg-post}), we choose examples in which the prior
 expectation is small ($E[b]=2$) and the number of observed events is
 small too ($k=0,1,\ldots,15$ counts), and find the 68.3\%, 90\%, and
 95\% posterior credible intervals for the signal, choosing central
 intervals if they contain the posterior mode or upper limits
 otherwise (both are invariant under reparametrization).  We consider
 uncertainties on the prior expectation of 10\%, 20\%, 50\%, and
 100\%, such that the shape and rate parameters of the background
 prior are listed in table~\ref{tab-bkg-priors-Eb2} and the
 corresponding prior densities are shown in
 figure~\ref{fig-bkg-priors-Eb2}.  The marginal model, for different
 choices of the background prior and sample size, is shown in
 figure~\ref{fig-marg-likelihoods}.  The corresponding marginal
 posteriors are shown in figure~\ref{fig-marg-posteriors}, and their
 numerical summaries are provided by the tables
 \ref{tab-post-summ-bkg-2-0.2}, \ref{tab-post-summ-bkg-2-0.4},
 \ref{tab-post-summ-bkg-2-1} and \ref{tab-post-summ-bkg-2-2} of
 appendix~\ref{app-tables-coverage}, in the form of left and right
 bounds of the 68.3\%, 90\%, 95\% credible intervals, mean, median,
 mode, variance, skewness and excess kurtosis.

 One aspect which deserves some comment is the case of zero observed
 counts.  Although the marginal posterior~(\ref{eq-marg-post}) does
 not explicitly depends on the expected background, it still contains
 it indirectly, due to the integration over the background prior.
 This implies that the upper limit in case of no counted events (as
 for any counts) does depend on the background prior.  This has been
 verified with a scan of the prior parameters, choosing a background
 expectation of $E[b] = 0.5, 1, 2, 4, 8$ counts and a relative
 uncertainty of 10\%, 20\%, 50\%, 100\%, and 150\%.  The results are
 shown in table~\ref{tab-bkg-UL95-zero-nObs}, which reports the
 posterior upper limit at 95\% credibility level for zero counts as a
 function of the prior background expectation and relative
 uncertainty.  The upper limit moves right with increasing background
 expectation, because the posterior becomes broader.  On the other
 hand, it has a small increase (negligible if $E[b]$ is very small)
 for increasing relative uncertainties up to 50\%, and then tends to
 decrease for increasing uncertainties, with a more pronounced trend
 for high background expectations.

 Quoting upper limits which depend on the prior background expectation
 in the case of zero observed counts produces a result which is
 similar to the classical results (summarized in \cite{pdb2010}) but
 may appear suboptimal from the Bayesian point of view.  The reason is
 that one important piece of information is ignored: in this case we
 know with certainty that there is no contribution from the background
 process, hence one can interpret the result using the simpler model
 in which the signal alone is described by a Poisson process.  Using
 the reference prior (which is Jeffreys' prior) in this case gives the
 reference posterior $p(s|0) = Ga(s; \frac{1}{2}, 1)$ and the upper
 limit, found as the 0.95-quantile of the Gamma density, is 1.92
 counts, which is lower than all values reported in
 table~\ref{tab-bkg-UL95-zero-nObs}.  One would obtain the same result
 with the model presented here, in the limit of a prior delta-function
 for the background prior which gives a unit mass to the single point
 $b=0$, because in this case one obtains the 1-dimensional Poisson
 model (by virtue of the result of appendix~\ref{sec-delta}).  This
 upper limit does not depend on the prior knowledge about the
 background and looks a bit ``aggressive'', compared to the well known
 upper limit of 3 counts which is obtained in the 1-dimensional case
 both using a flat prior (which is falsely considered non-informative
 and leads to solutions which are not invariant under
 reparametrization) and the classical approach \cite{pdb2010}.  On the
 other hand, it is the result of including all available information
 about the problem under consideration.

\begin{table}[h]
\centering
\begin{tabular}{r|ccccc}
$E[b]$   & \multicolumn{5}{c}{Relative uncertainty on $E[b]$:} \\
$\downarrow$\;~   & 10\% & 20\% & 50\% & 100\%   &   150\%  \\
\hline
0.5   &   2.55   &   2.55   &   2.56   &   2.57   &   2.55   \\
1.0   &   2.65   &   2.66   &   2.68   &   2.66   &   2.60   \\
2.0   &   2.77   &   2.78   &   2.80   &   2.74   &   2.62   \\
4.0   &   2.88   &   2.89   &   2.91   &   2.78   &   2.62   \\
8.0   &   2.96   &   2.98   &   2.99   &   2.81   &   2.62   \\
\end{tabular}
\caption{Upper limits at 95\% credibility level for zero observed counts.}
\label{tab-bkg-UL95-zero-nObs}
\end{table}

 Given the observed counts and the prior beliefs on the background
 values, the method described in this paper provides an objective
 Bayesian solution for the statistical inference about the true
 signal, the marginal reference posterior~(\ref{eq-marg-post}).  In
 the context of a specific analysis, it is important to study the
 sampling properties of the solution.  In the large sample size limit,
 Bayesian $q$-credible regions are always approximate confidence
 regions with coverage $q$ and in some case they have even the exact
 coverage, as it happens for location-scale models
 \cite{Bernardo2007}.  However, the actual measurement may be far from
 the asymptotic regime.  In addition our model is discrete, such that
 the coverage is not exact (of course this is also true for
 frequentist solutions), even though the fluctuations around the
 coverage $q$ become smaller and smaller with increasing sample sizes.

 A frequentist study of the coverage for different solutions of the
 Poisson problem, in which the reference posterior (which corresponds
 to the use of the Jeffreys' prior) is also included, is presented in
 Ref.~\cite{Cousins2010}.  We adopted a similar treatment to study few
 examples in which the small sample size is clearly far from the
 asymptotic limit, such that the coverage properties are not expected
 to be ideal.  Different possible observations ($k=0,1,\ldots,15$
 counts) have been considered.  The coverage properties are a function
 of five parameters: the true signal and background yields, the shape
 and rate parameters of the background prior, and the observed count.
 The first step is to average over the possible observations with the
 corresponding Poisson weights, which leaves four degrees of freedom.
 We report the results obtained with the four background priors which
 correspond to an expectation of $E[b]=2$ with relative uncertainty of
 10\%, 20\%, 50\%, and 100\% (the actual study considered background
 expectations of 0.5, 1, 2, 4, and 8 counts and also included the case
 of 5\% and 150\% relative uncertainties, but the qualitative features
 are the same as the examples presented here).  For each prior, we
 still have two degrees of freedom left, the true signal and
 background yields.  Hence the coverage can be shown in the form of
 2-dimensional histograms.

 Figures \ref{fig-coverage-Eb2-0.2}, \ref{fig-coverage-Eb2-0.4},
 \ref{fig-coverage-Eb2-1} and \ref{fig-coverage-Eb2-2} in
 appendix~\ref{app-tables-coverage} show the coverage of 68.3\%, 90\%
 and 95\% posterior credible intervals as a function of the true
 signal and background values.  The diagonal structure is clearly due
 to the fact that we observe a single quantity, the number of counted
 events, which is the best estimator of the sum of true signal and
 background.  This pattern is illustrated by the diagonal dotted
 lines, which characterize the loci with constant (and integer) sum of
 signal and background but different signal fraction.  These figures
 also show the coverage as a function of the true signal alone, in the
 assumption that the true background coincides with the prior
 expectation (the 1D plot is the slice of the 2D plot at its left
 along the horizontal dashed line).  As expected, a wider prior
 uncertainty gives more conservative results, in the sense that the
 overall tendency is to overcover, apart from very small values of the
 true signal, for which there is undercoverage.  Instead, a narrow
 prior leads to a more symmetrical distribution of the actual coverage
 about the nominal one (the horizontal line in the right plots).

 It is interesting to note what happens if the true signal is quite
 different from the prior expectation.  In
 figure~\ref{fig-coverage-Eb2-2-b1e3}
 (appendix~\ref{app-tables-coverage}) the coverage of the solution
 which corresponds to a prior expectation of $E[b]=2$ with 100\%
 uncertainty is shown when the true background is half of it (left
 plots) or 50\% bigger (right plots).  The tendency to overcover is
 more pronounced when the background is smaller than expected (which
 comes at no surprise).  When the true background is half standard
 deviation higher than expected, the pattern is the same as for the
 case in which it matches the expectation (right plots of
 figure~\ref{fig-coverage-Eb2-2}) but the fluctuations about the
 nominal coverage are less pronounced.

 Other figures of merit for the solution are the false exclusion rate
 in case of 95\% upper limits and the bias of the signal estimators.
 In appendix~\ref{app-tables-coverage} the case of prior background
 expectation $E[b]=2$ counts with relative uncertainties of 10\%, 50\%
 and 100\% are shown (figures \ref{fig-FER-biasP-Eb2-Sb10},
 \ref{fig-biasE-biasM-Eb2-Sb10}, \ref{fig-FER-biasP-Eb2-Sb50},
 \ref{fig-biasE-biasM-Eb2-Sb50}, \ref{fig-FER-biasP-Eb2-Sb100}).  The
 diagonal pattern is still visible in the plots of the false exclusion
 rate as a function of the true signal and background, but not in the
 bias plots.  The false exclusion rate may be larger than 5\% when the
 prior uncertainty is small but this never happens when it is 100\% or
 larger and it is very rare with 50\% uncertainty.  The bias of the
 different estimators, in all cases, is a monotonically decreasing
 function of the true signal.  The mode tends to underestimate the
 true signal when the prior uncertainty is small, although bigger
 prior uncertainties tend to produce a more symmetric behaviour.  The
 difference with respect to the true signal is usually less than one
 unit.  On the other hand, both the mean and median tend to
 overestimate the true signal.  Although the mean appears a bit more
 balanced than the mean, with a larger difference with respect to the
 true signal than the bias of the mode.  In conclusion, the intuitive
 choice of the posterior peak as the best estimator for the true
 signal comes out to be the least biased estimator for small sample
 sizes.


\section{Summary and conclusion}\label{sec-conclusion}

 We considered the signal+background model in which both sources are
 independent Poisson processes, and there is available prior
 information about the background intensity $b$ encoded in a Gamma
 density of the form~(\ref{eq-gamma}). Following the prescription by
 \cite{Sun1998}, the reference prior for the signal parameter $s$ is
 computed from the conditional model~(\ref{eq-marg-mod-2}).  The
 reference prior is proportional to the square root of the Fisher's
 information~(\ref{eq-fisher-info-2}), and is not normalizable.  This
 is not a problem, because the reference posterior is a proper
 density, and the same also happens when using a $\delta$-function as
 the prior for the background.  In this case the Jeffreys' prior
 $\pi(s) \propto (s+b_{0})^{-1/2}$ is an improper density too, which
 cannot represent somebody's degree of belief.  The justification for
 the use of an improper prior is provided in the framework of the
 Bayesian reference analysis, in which the reference prior is defined
 as the mathematical device which maximizes the amount of missing
 information and does not represent an actual degree of belief
 \cite{Bernardo2005a}.

 Being an improper prior, one can always scale it by a constant: the
 normalization of the posterior density can be computed once the
 latter is known.  For practical applications, we propose to adopt the
 improper prior~(\ref{eq-ref-prior-signal}) obtained by dividing the
 square root of the Fisher's information by the value which it assumes
 for $s=0$.  The resulting function assumes its maximum value (equal
 to one) at the origin and uniformly decreases for increasing values
 of $s$.  Of course, one is also free to use
 equation~(\ref{eq-sqrt-fisher-info}) directly or to scale it by any
 other constant value.

 The joint prior for our model is the improper function $\pi(s,b) =
 \pi(s) \, p(b)$ where $\pi(s)$ is defined by
 equation~(\ref{eq-ref-prior-signal}) (or any other function which is
 proportional to $|I(s)|^{1/2}$) and $p(b)$ is the Gamma
 density~(\ref{eq-gamma}) which encodes the experimenter's prior
 degree of belief about the background.  Finally, the marginal
 reference posterior is proportional to the product of the reference
 prior (\ref{eq-ref-prior-signal}) and the marginal likelihood
 (\ref{eq-marg-mod-2}):
\[
  p(s|k) \propto \left( \frac{\beta}{1+\beta} \right)^{\!\alpha}
                 e^{-s} \, f(s;k,\alpha,\beta) \, \pi(s) \; .
\]
 A few examples of the application of this solution have been shown in
 section~\ref{sec-posterior}, together with their coverage, false
 exclusion rate, and the bias of different estimators of the true signal.

\section*{Acknowledgments}

 The author thanks Jim Berger, Rostislav Konoplich, Sven Kreiss and
 Kyle Cranmer for the careful reading of this draft and the useful
 suggestions.



%

\appendix

\section{Computing the conditional model}\label{sec-code}

 All results shown in this paper can be easily implemented once the
 function 
\begin{equation*}
  f(s;k,\alpha,\beta) = \sum_{n=0}^{k} \binom{\alpha+n-1}{n}
             \frac{s^{k-n}}{(k-n)! \, (1+\beta)^{n}}
\end{equation*}
 is available.  Here we give some suggestion which lead to a fast
 evaluation of $f(s;k,\alpha,\beta)$.

 Let us focus on each addendum first.  It is better to work with the
 logarithms, because this avoids rounding problems related to
 expressions featuring very big and very small values:
\begin{equation*}
\begin{split}
  &\binom{\alpha+n-1}{n} \frac{s^{k-n}}{(k-n)! \, (1+\beta)^{n}} =
\\
  &= \exp\left[ \log\binom{\alpha+n-1}{n} + (k-n) \log s \right.
\\
  &\left. \quad - n \log(1+\beta) - \log(k-n)!
     \right]
\end{split}
\end{equation*}
 for $s>0$.  For $s=0$ one computes directly the constant term which
 corresponds to $n=k$ in the sum from equation~(\ref{eq-f-zero}).

 Plotting or integrating $f(s;k,\alpha,\beta)$ as a function of $s$,
 possibly for different values of $k$, implies that $\alpha,\beta$ are
 held constant during the computation.  We now consider all terms from
 left to right.

 The binomial coefficient is a function of $n$ with $\alpha$ fixed
 which increases faster than $\alpha^n$ and should be tabulated during
 the initialization phase.  The following recursive relation holds:
\begin{equation*}
  \binom{\alpha+n-1}{n} =
\begin{cases}
  1 &\text{if} \; n=0 \\
  \alpha &\text{if} \; n=1 \\
  \displaystyle
  \frac{\alpha+n-1}{n} \, \binom{\alpha+n-2}{n-1} &\text{if} \; n\ge2 \\
\end{cases}
\end{equation*}
 from which it follows that $a(n;\alpha) \equiv \log
 \binom{\alpha+n-1}{n} = \log[1+(\alpha-1)/n] + a(n-1;\alpha)$.  For
 example, one can start by saving into a C++ vector the $a(n;\alpha)$
 values for $n=0,\ldots,100$ and add more terms if needed later on,
 such that computing it during the loop over $n$ reduces to accessing
 the $n$-th element of the vector.

 The value of $\log s$ should be computed once, outside the sum.

 Because $\beta$ is fixed, $\log(1+\beta)$ is a constant term which
 should be only computed during the initialization phase, such that
 $b(n;\beta) \equiv n\log(1+\beta) = n \, b(1;\beta)$ is just a
 multiplication step.  For intensive computations (like summations
 over $k=0,\ldots,\infty$ and/or integrations over $s\in[0,\infty[$)
 it can be useful to also store these values during the initialization
 phase.

 Finally, the term $c(m) \equiv \log m! = \log m + c(m-1)$, where
 $m=k-n$, should be also computed during the initialization phase (say
 for $m=0,\ldots,100$) and stored into a vector for direct access
 inside the loop.


\section{Properties of $f(s;k,\alpha,\beta)$: Proofs}\label{sec-proofs}

\subsection*{Proof of property 1}

 Property 3 is equivalent to the statement that $E[1]=1$,
 which is explicitly shown here:
\begin{equation*}
\begin{split}
  E[1] &= \sum_{k=0}^{\infty} p(k|s)
         = \sum_{k=0}^{\infty} \left( \frac{\beta}{1+\beta} \right)^{\!\alpha}
           e^{-s} \, f(s;k,\alpha,\beta)
   \\
       &= \left( \frac{\beta}{1+\beta} \right)^{\!\alpha} e^{-s} \,
          \sum_{k=0}^{\infty} \sum_{n=0}^{k}
          \frac{s^{k-n}}{(k-n)!} \binom{\alpha+n-1}{n}
          \frac{1}{(1+\beta)^n}
   \\
       &= \left( \frac{\beta}{1+\beta} \right)^{\!\alpha} e^{-s} \,
          \sum_{n=0}^{\infty}
          \binom{\alpha+n-1}{n} \frac{1}{(1+\beta)^n}
    \sum_{k=n}^{\infty} \frac{s^{k-n}}{(k-n)!}
          \quad\; \left\{ \sum_{k=0}^{\infty} \sum_{n=0}^{k} =
              \sum_{n=0}^{\infty} \sum_{n=k}^{\infty} \right\}
   \\
       &= \left( \frac{\beta}{1+\beta} \right)^{\!\alpha} e^{-s} \,
          \sum_{n=0}^{\infty}
          \binom{\alpha+n-1}{n} \frac{1}{(1+\beta)^n}
    \sum_{m=0}^{\infty} \frac{s^{m}}{m!}
          \qquad \qquad \{ m \equiv k-n \}
   \\
       &= \left( \frac{\beta}{1+\beta} \right)^{\!\alpha} 
          \sum_{n=0}^{\infty}
          \binom{\alpha+n-1}{n} \left(\frac{1}{1+\beta}\right)^{\!n}
   \\
       &= \left( \frac{\beta}{1+\beta} \right)^{\!\alpha} 
          \frac{1}{[1 - 1/(1+\beta)]^{\alpha}}
        = 1 
\end{split}
\end{equation*}
 where the following theorems have been used:
\[
  \sum_{m=0}^{\infty} \frac{s^{m}}{m!} = e^s
  \qquad \text{and} \qquad
  \sum_{n=0}^{\infty} \binom{\alpha+n-1}{n} x^{n} = \frac{1}{(1-x)^\alpha}
  \; .
\]


\subsection*{Proof of property 2}

 We proceed by induction, showing that
 equation~(\ref{eq-f-derivatives}) holds for $n=1$ and that if it
 holds for $n-1$ then it must also hold for $n$.  For $n=1$, by direct
 computation one obtains
\begin{equation*}
\begin{split}
  f^{(1)}(s;k,\alpha,\beta) &= \sum_{n=0}^{k-1} \frac{s^{k-n-1}}{(k-n-1)!}
                         \binom{\alpha+n-1}{n}
       \frac{1}{(1+\beta)^n}
       \\
 &=
\begin{cases}
  0                      &\text{if}\quad k=0 \\
  1                      &\text{if}\quad k=1 \\
  f(s;k-1,\alpha,\beta)  &\text{if}\quad k\ge1 \\
\end{cases}
\end{split}
\end{equation*}

 The next step is to show that if the property
 (\ref{eq-f-derivatives}) holds for $n-1$ then it must also hold for
 $n$.  For $n\le k$ this is trivial (because $f^{(n-1)}$ is zero),
 while for $n>k$
\[
\begin{split}
  \frac{\partial^n}{\partial s^n} f(s;k,\alpha,\beta) &=
  \frac{\partial}{\partial s} \frac{\partial^{n-1}}{\partial s^{n-1}}
                              f(s;k,\alpha,\beta)
            \\
  &= \frac{\partial}{\partial s} f(s;k-n+1,\alpha,\beta)
            \\
  &= f(s;k-n,\alpha,\beta)
\end{split}
\]
 where the last passage comes from $f^{(1)}(s;m,\alpha,\beta) =
 f(s;m-1,\alpha,\beta)$, in which $m=k-n+1$.  This completes the proof.




\subsection*{Proof of property 3}

 Property 3 follows directly from the previous ones:
\begin{equation}
\begin{split}
  E\left[
    \frac{f^{(n)}(s;k,\alpha,\beta)}{f(s;k,\alpha,\beta)}
    \right]
   &= \sum_{k=0}^{\infty} \frac{f^{(n)}}{f} \, p(k|s)
    = \sum_{k=0}^{\infty} \frac{f^{(n)}}{f}
     \left( \frac{\beta}{1+\beta} \right)^{\!\alpha} e^{-s} \, f
 \\
   &= \sum_{k=n}^{\infty}
      \left( \frac{\beta}{1+\beta} \right)^{\!\alpha} e^{-s} \, 
       f^{(n)}(s;k,\alpha,\beta)
      \qquad \{ k<n \Rightarrow f^{(n)}=0 \}
 \\
   &= \sum_{m=0}^{\infty}
      \left( \frac{\beta}{1+\beta} \right)^{\!\alpha} e^{-s} \, 
       f(s;m,\alpha,\beta)
      \qquad \qquad \{ m \equiv k-n \}
 \\
   &= E[1] = 1
\end{split}
\end{equation}


\section{Comparison with the DJP and PPSS papers}\label{sec-other-papers}

 The notation used in the DJP and PPSS papers is different from the
 one adopted here:
\begin{center}
\begin{tabular}{ccc|ccc}
 DJP           & PPSS  & here & DJP & PPSS & here \\
\hline
 $\sigma$      & $s$   & $s$  & $b$ & $b$  & $\beta$ \\
 $\mu$         & $\mu$ & $b$  & $y$ & $Y$  & $\alpha+\frac{1}{2}$ \\
 $\varepsilon$ & 1     & 1    & $n$ & $N$  & $k$ \\
\multicolumn{3}{c|}{--- model--- } &
\multicolumn{3}{c}{--- bkg prior ---}
\end{tabular}
\end{center}

 The DJP model has one more parameter which is taken to be equal to
 1 both here and in the PPSS paper.  DJP wrote the Gamma density
 which represents such parameter in this form:
\[
  \pi(\varepsilon) = \frac{a(a\varepsilon)^{x-1/2} \, e^{-a\varepsilon}}
                          {\Gamma(x+1/2)}
\]
 By formally setting $a=x$ and taking the limit $x\to\infty$ this
 distribution degenerates in the delta-function
 $\delta(\varepsilon-1)$, which is what is considered here and by
 PPSS.

 The marginal model of DJP is
\[
  p(n|\sigma) = \left(\frac{a}{a+\sigma}\right)^{\!x+1/2}
                \left(\frac{b}{b+1}\right)^{\!y+1/2}
                S_{n}^{0}(\sigma)
\]
 where one polynomial appears of the following family
\[
  S_{n}^{m}(\sigma) = \sum_{k=0}^{n} \binom{k+x-\frac{1}{2}}{k}
                      \binom{n-k+y-\frac{1}{2}}{n-k}
                      \left(\frac{\sigma}{a+\sigma}\right)^{\!k}
                      \frac{k^m}{(b+1)^{n-k}}
\]
 In the limit $a=x\to\infty$ the polynomials above become
\[
  S_{n}^{m}(\sigma) \to
  \sum_{k=0}^{n} \binom{n-k+y-\frac{1}{2}}{n-k}
  \frac{k^m \, \sigma^k}{k! \, (b+1)^{n-k}}
\]
 which in the notation of the PPSS paper coincides with
\(
  e^s \, T_{n}^{m}(s)
\).

 In the same limit, the DJP marginal model becomes
\[
  p(n|\sigma) \to e^{-\sigma} \left(\frac{b}{b+1}\right)^{\!y+1/2}
                  \sum_{k=0}^{n} \binom{n-k+y-\frac{1}{2}}{n-k}
                  \frac{\sigma^k}{k! \, (b+1)^{n-k}}       
\]
 In the notation of this paper, the expression above becomes
\[
\begin{split}
 p(n|\sigma) &= 
   e^{-s} \left(\frac{\beta}{\beta+1}\right)^{\!\alpha}
   \sum_{k=0}^{n} \binom{n-k+\alpha-1}{n-k}
   \frac{s^k}{k! \, (\beta+1)^{n-k}} =
\\
   \left\{ \text{set $m \equiv n-k$ in the sum} \right\}
\\
   &= e^{-s} \left(\frac{\beta}{\beta+1}\right)^{\!\alpha}
   \sum_{m=0}^{n} \binom{m+\alpha-1}{m}
   \frac{s^{n-m}}{(n-m)! \, (\beta+1)^{m}} =
\\
   &= e^{-s} \left(\frac{\beta}{\beta+1}\right)^{\!\alpha}
   f(s;n,\alpha,\beta)  \quad \leftarrow
   \text{eq.~(\ref{eq-marg-mod-2}) of section~\ref{sec-marginal-model}.}
\end{split}
\]

 In a similar way, one can write the marginal model in the PPSS paper
 in the form of our equation~(\ref{eq-marg-mod-2}), because the
 function $e^{-s}\,f(s;n,\alpha,\beta)$ is the same thing as the PPSS
 function $T_n^0(s)$.


\section{Model with a degenerate background prior}\label{sec-delta}

 If we have a certain knowledge about the background yield, the prior
 for $b$ becomes a delta-function $p(b) = \delta(b-b_{0})$, which is
 the degenerate form of a Gamma density with parameters
 $\beta=\alpha/b_{0}, \alpha\to\infty$.  In this case, the marginal
 model is
\[
  p(k|s) = \int_0^\infty \text{Poi}(k|s+b) \, \delta(b-b_{0}) \di b
         = \text{Poi}(k|s+b_{0})
\]
 whose logarithm is
\[
  \log p(k|s) = k \log (s+b_{0}) -s - b_{0} - \log k! \; .
\]
 The second derivative of the logarithm of the marginal model is
 $-k/(s+b_{0})^2$ such that the Fisher's information is
\[
\begin{split}
  I(s) &= - E\left[ \left( \frac{\partial^2}{\partial s^2}
            \log p(k|s) \right) \right] 
\\
       &= e^{-s - b_{0}} \, \sum_{k=0}^{\infty}
          \frac{k}{(s+b_{0})^2} \frac{(s+b_{0})^k}{k!}
\end{split}
\]
 whose 0-th element is null, hence we can write a sum starting from
 $k=1$ and then introduce a new index $n=k-1$:
\[
\begin{split}
  I(s) &= e^{-s - b_{0}} \, \sum_{k=1}^{\infty}
          \frac{(s+b_{0})^{k-2}}{(k-1)!}
\\
       &= \frac{e^{-s - b_{0}}}{s+b_{0}} \, \sum_{n=0}^{\infty}
          \frac{(s+b_{0})^n}{n!}
        = \frac{1}{s+b_{0}}
\end{split}
\]
 from which one finds the reference prior as $\pi(s)
 \propto |I(s)|^{1/2} = (s+b_{0})^{-1/2}$.

 We now verify that one gets the same solution when defining
 $\beta=\alpha/b_{0}$ and considering the limit $\alpha\to\infty$ in
 our model.  First, let's consider the function $f(s;k,\alpha,\beta)$:
\[
\begin{split}
  f(s;k,\alpha,\alpha/b_{0}) &= \sum_{n=0}^{k}
   \frac{\Gamma(\alpha+n)}{\Gamma(\alpha)}
   \frac{s^{k-n}}{(k-n)!}
   \frac{b_{0}^{n}}{n!\,(b_{0}+\alpha)^{n}}
\\
    & \xrightarrow{\alpha\to\infty} \sum_{n=0}^{k}
      \frac{s^{k-n} \, b_{0}^{n}}{(k-n)! \, n!}
    = \frac{(s+b_0)^k}{k!}  \; .
\end{split}
\]
 Let's call this limit $f(s;k) = (s+b_0)^k / k!$.  Next, consider
\[
\begin{split}
  \left(\frac{\beta}{1+\beta}\right)^{\!\alpha}
     &= \left(\frac{\alpha}{\alpha+b_0}\right)^{\!\alpha}
\\
     &= \left(1+\frac{b_0}{\alpha}\right)^{\!-\alpha}
     \xrightarrow{\alpha\to\infty} e^{-b_0}  \; .
\end{split}
\]

 Finally, in this limit the Fisher's information
 (\ref{eq-sqrt-fisher-info}) becomes
\[
\begin{split}
  I(s) &= e^{-s-b_0} \, \sum_{n=0}^{\infty}
          \frac{f^2(s;n)}{f(s;n+1)} - 1
\\
       &= e^{-s-b_0} \, \sum_{n=0}^{\infty}
          \left[\frac{(s+b_0)^n}{n!}\right]^2
          \frac{(n+1)!}{(s+b_0)^{n+1}} - 1
\end{split}
\]
 The 0-th term in the sum is $(s+b_0)^{-1}$ and, from the 1-st term
 on, the factorial $(n-1)!$ is well defined, such that we can write
\[
  I(s) = \frac{e^{-s-b_0}}{s+b_0} + e^{-s-b_0} \,
         \left[ \sum_{n=1}^{\infty}
           \frac{(s+b_0)^{n-1}}{(n-1)!} \frac{n+1}{n}
         \right] - 1
\]

 \vspace{4em}

 The sum above can be rewritten
\[
\begin{split}
  \sum_{n=1}^{\infty} & \frac{(s+b_0)^{n-1}}{(n-1)!} \frac{n+1}{n}
      = \sum_{m=0}^{\infty} \frac{(s+b_0)^{m}}{m!}
            \left( 1 + \frac{1}{m+1} \right)
  \\
      &= e^{s+b_0} + \sum_{m=0}^{\infty} \frac{(s+b_0)^{m}}{m! \, (m+1)}
  \\
      &= e^{s+b_0} + (s+b_0)^{-1} 
             \sum_{m=0}^{\infty} \frac{(s+b_0)^{m+1}}{(m+1)!}
  \\
      &= e^{s+b_0} + (s+b_0)^{-1} 
             \sum_{n=1}^{\infty} \frac{(s+b_0)^{n}}{n!}
  \\
      &= e^{s+b_0} + \frac{e^{s+b_0} - 1}{s+b_0}
\end{split}
\]
 which inserted in $I(s)$ gives
\[
\begin{split}
  I(s) &= \frac{e^{-s-b_0}}{s+b_0} + e^{-s-b_0} \,
         \left[ e^{s+b_0} + \frac{e^{s+b_0} - 1}{s+b_0}
         \right] - 1
  \\
       &= \frac{1}{s+b_0}  \; ,
\end{split}
\]
 exactly the same expression which one obtains when starting directly
 with the delta-function.


\section{Summaries of the solutions and their properties}\label{app-tables-coverage}

 The following tables and plots provide a summary of the marginal
 posteriors for the signal in the cases of $E[b]=2$ prior background
 expectation with relative uncertainty of 10\%, 20\%, 50\% and 100\%,
 and of their coverage.  In addition, the false exclusion rate and
 biases of the posterior mode, mean and median are shown.

\begin{figure*}[t!]
 \centering
 \includegraphics[width=\textwidth]{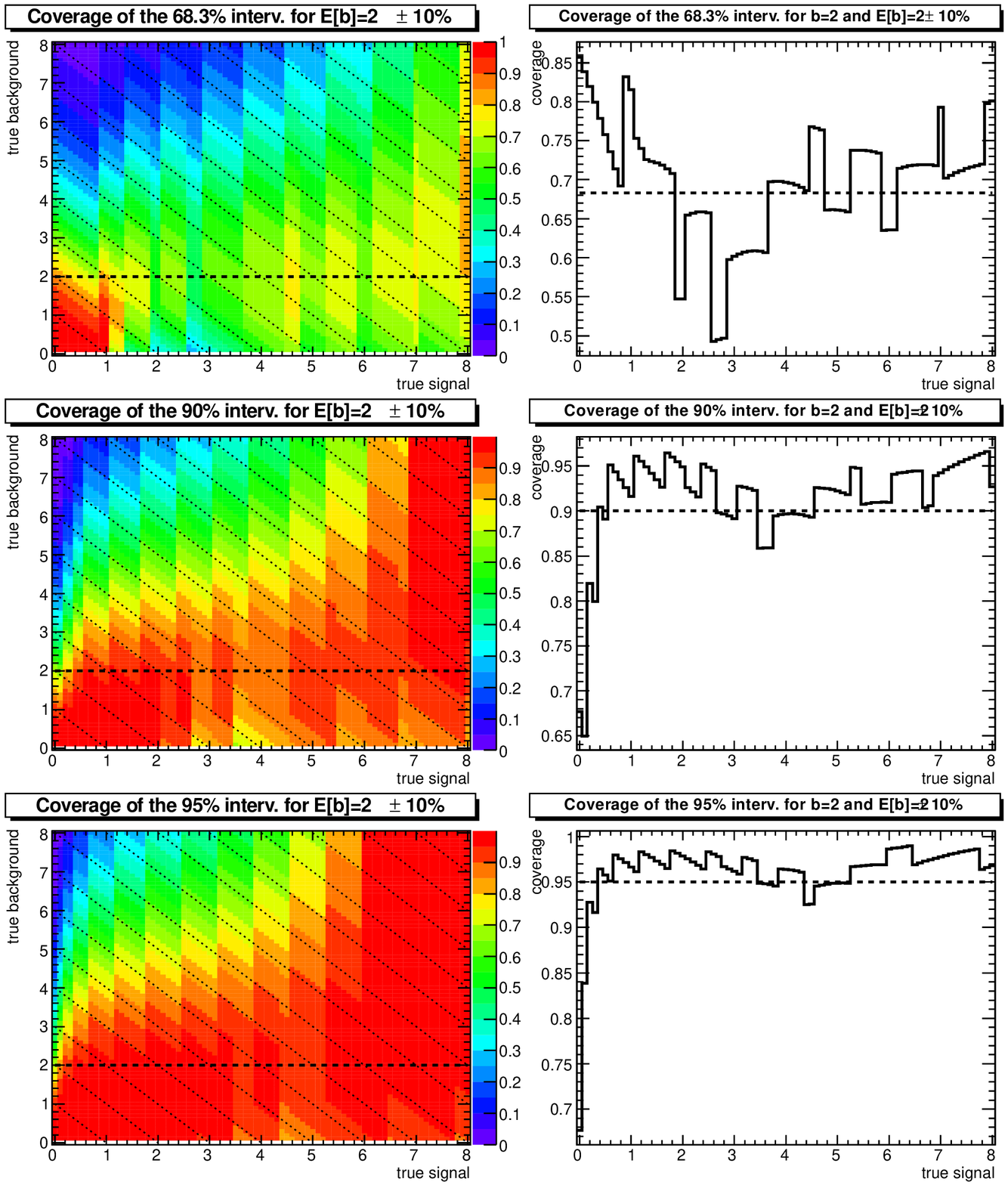}
 \caption{Coverage of the 95\% (top), 90\% (middle), 68.3\% (bottom)
 posterior credible intervals for $E[b] = 2 \pm 0.2$ as a function of
 the true signal and background (left panel) and as a function of the
 true signal alone in case the true background coincides with the
 prior expectation (left panel).}
 \label{fig-coverage-Eb2-0.2}
\end{figure*}

\begin{figure*}[t!]
 \centering
 \includegraphics[width=\textwidth]{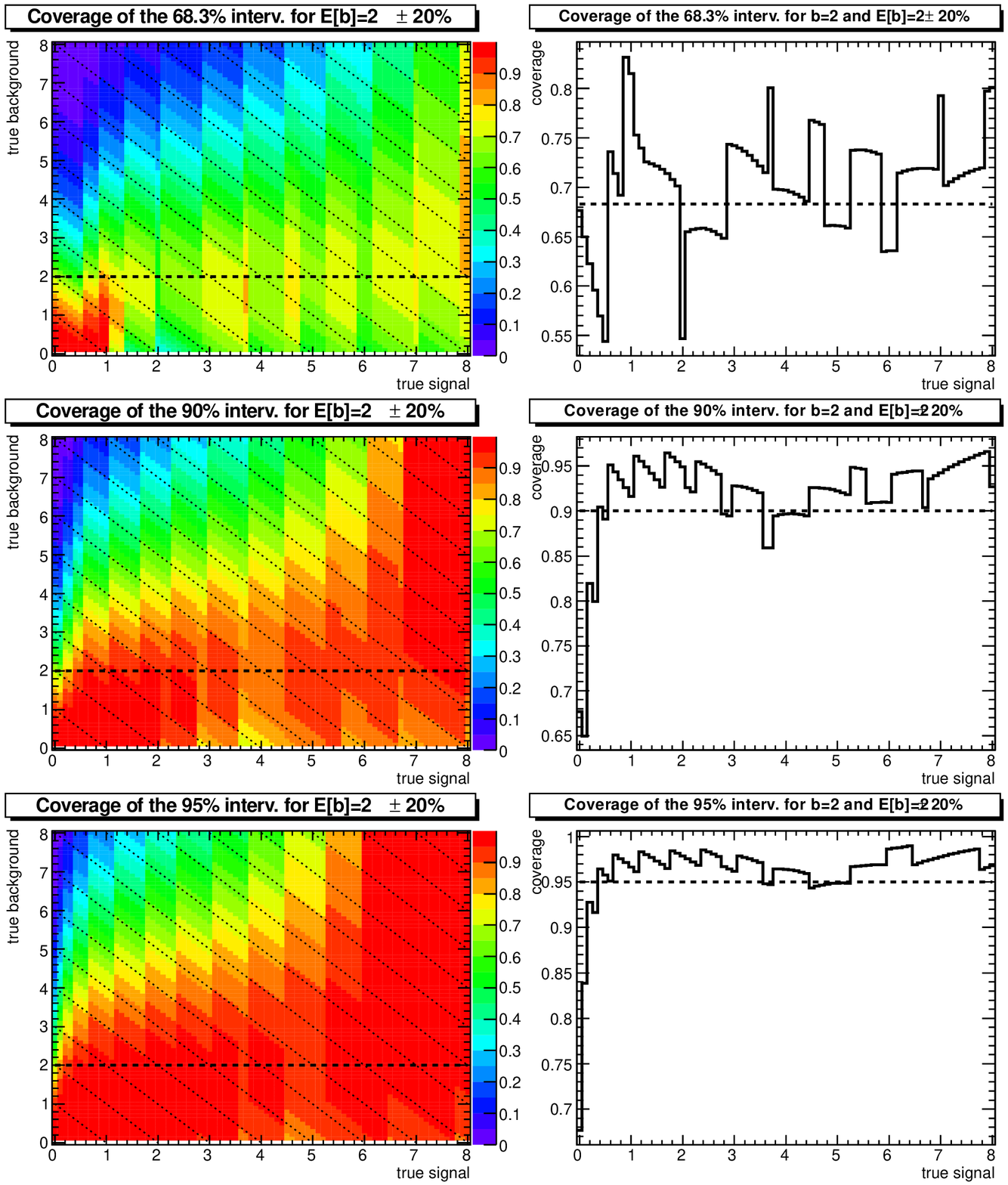}
 \caption{Coverage of the 95\% (top), 90\% (middle), 68.3\% (bottom)
 posterior credible intervals for $E[b] = 2 \pm 0.4$ as a function of
 the true signal and background (left panel) and as a function of the
 true signal alone in case the true background coincides with the
 prior expectation (left panel).}
 \label{fig-coverage-Eb2-0.4}
\end{figure*}

\begin{figure*}[t!]
 \centering
 \includegraphics[width=\textwidth]{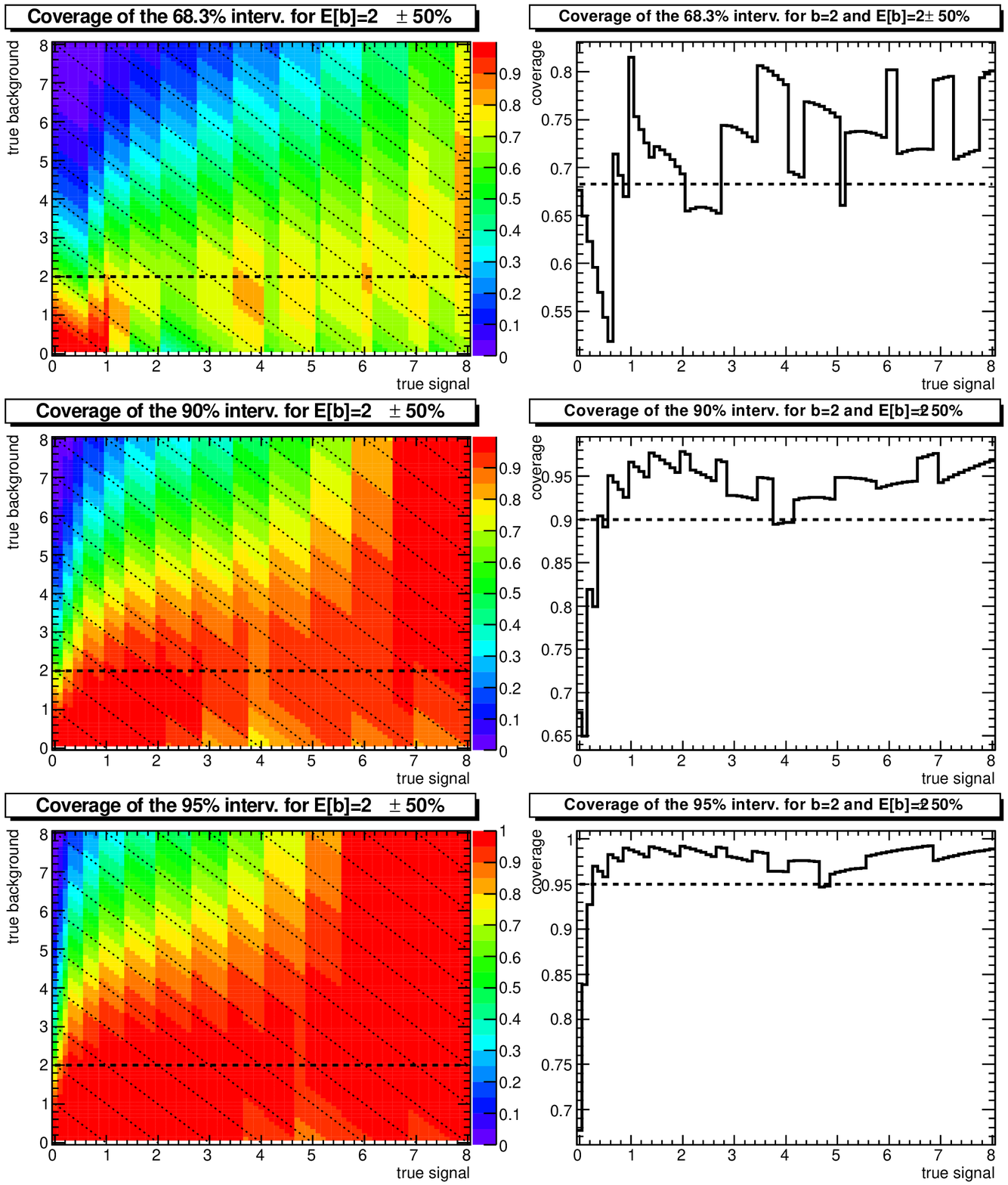}
 \caption{Coverage of the 95\% (top), 90\% (middle), 68.3\% (bottom)
 posterior credible intervals for $E[b] = 2 \pm 1$ as a function of
 the true signal and background (left panel) and as a function of the
 true signal alone in case the true background coincides with the
 prior expectation (left panel).}
 \label{fig-coverage-Eb2-1}
\end{figure*}

\begin{figure*}[t!]
 \centering
 \includegraphics[width=\textwidth]{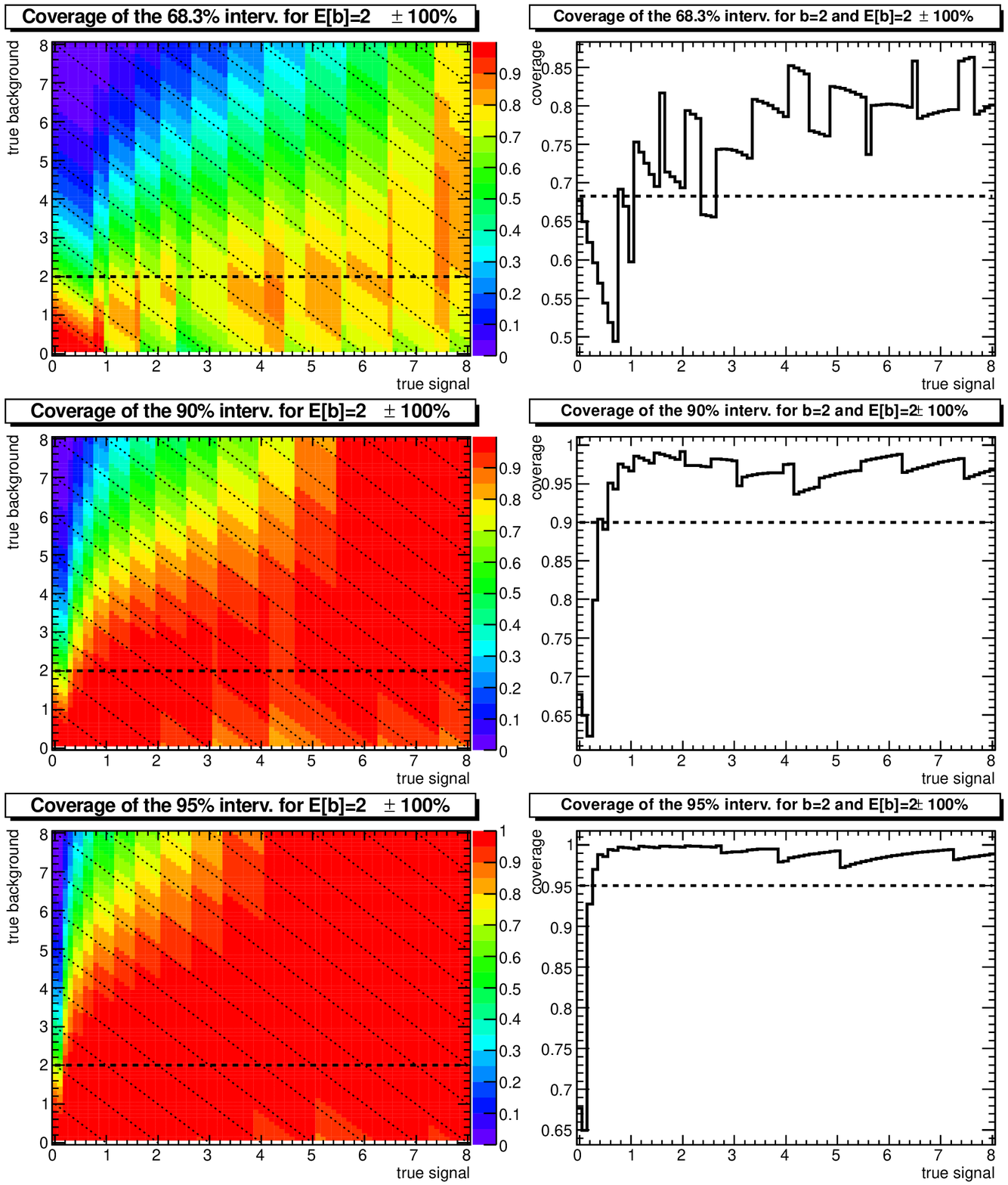}
 \caption{Coverage of the 95\% (top), 90\% (middle), 68.3\% (bottom)
 posterior credible intervals for $E[b] = 2 \pm 2$ as a function of
 the true signal and background (left panel) and as a function of the
 true signal alone in case the true background coincides with the
 prior expectation (left panel).}
 \label{fig-coverage-Eb2-2}
\end{figure*}

 Tables \ref{tab-post-summ-bkg-2-0.2}, \ref{tab-post-summ-bkg-2-0.4},
 \ref{tab-post-summ-bkg-2-1} and \ref{tab-post-summ-bkg-2-2} report
 the left and right bounds of the 68.3\%, 90\%, 95\% credible
 intervals, plus mean, median, mode, variance, skewness and excess
 kurtosis.  The coverage of such intervals has been studied for
 different observations ($k=0,1,\ldots,15$ counts) and the average
 over $k$ is shown in figures \ref{fig-coverage-Eb2-0.2},
 \ref{fig-coverage-Eb2-0.4}, \ref{fig-coverage-Eb2-1} and
 \ref{fig-coverage-Eb2-2}.  Because of the limited range in $k$
 considered here, the coverage is to be intended as a first
 approximation when the sum of true signal and background exceeds
 10--12 counts.  In the 2D plots, the diagonal lines have constant sum
 of true signal and background, whereas the horizontal line identifies
 the slice which is shown in the corresponding 1D plot.  The
 horizontal line in 1D plots shows the nominal coverage.

\begin{figure*}[t!]q
 \centering
 \includegraphics[width=\textwidth]{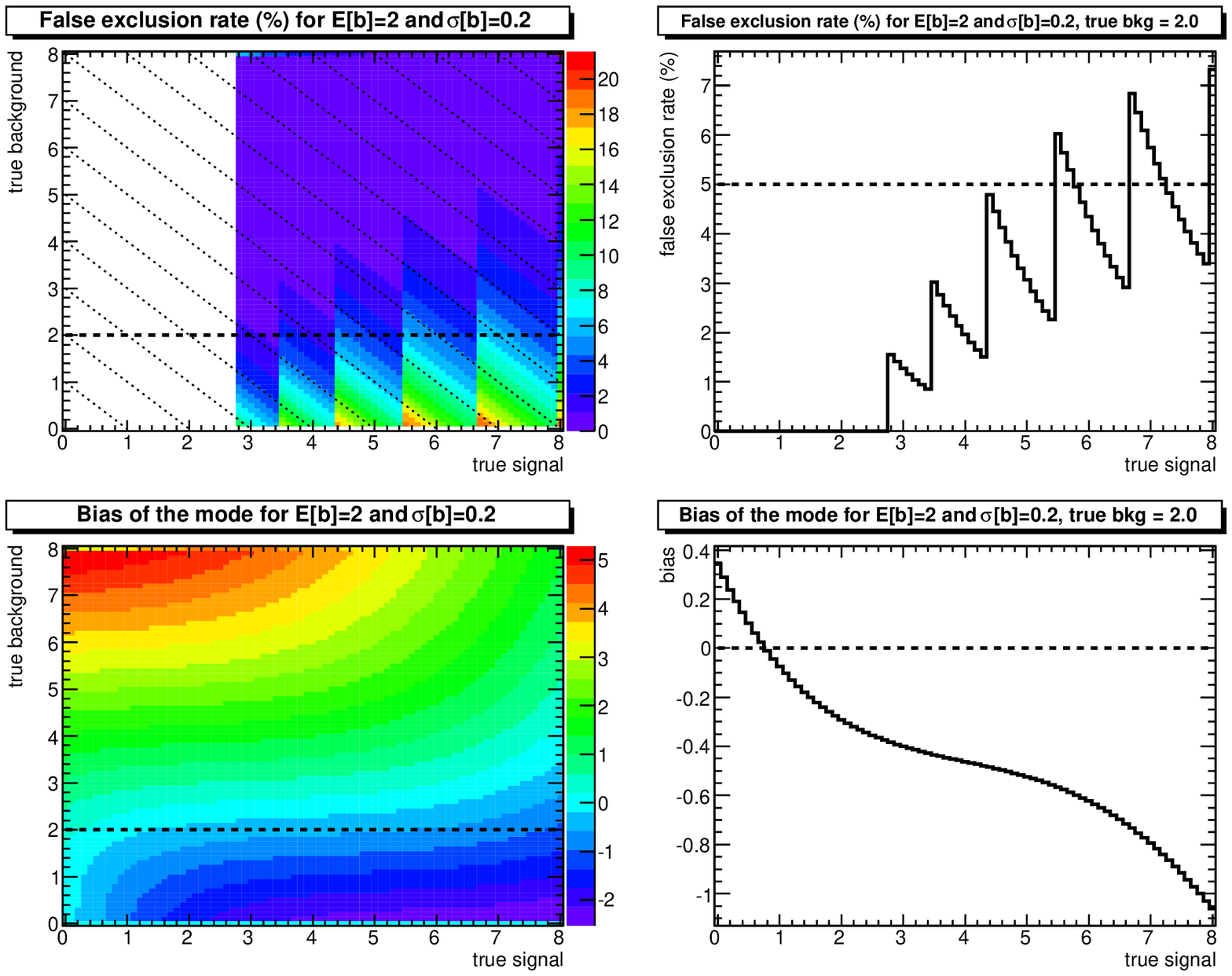}
 \caption{Top: false exclusion rate as a function of the true signal
   and background (top-left) and of the true signal alone for the case
   in which the true background coincides with the prior expectation
   (top-right).  Bottom: bias of the posterior mode as a function of
   the true signal and background (bottom-left) and of the true signal
   alone for the case in which the true background coincides with the
   prior expectation (bottom-right).  The posterior corresponds to a
   prior background expectation of 2 counts with 10\% relative
   uncertainty.}
 \label{fig-FER-biasP-Eb2-Sb10}
\end{figure*}

\begin{figure*}[t!]
 \centering
 \includegraphics[width=\textwidth]{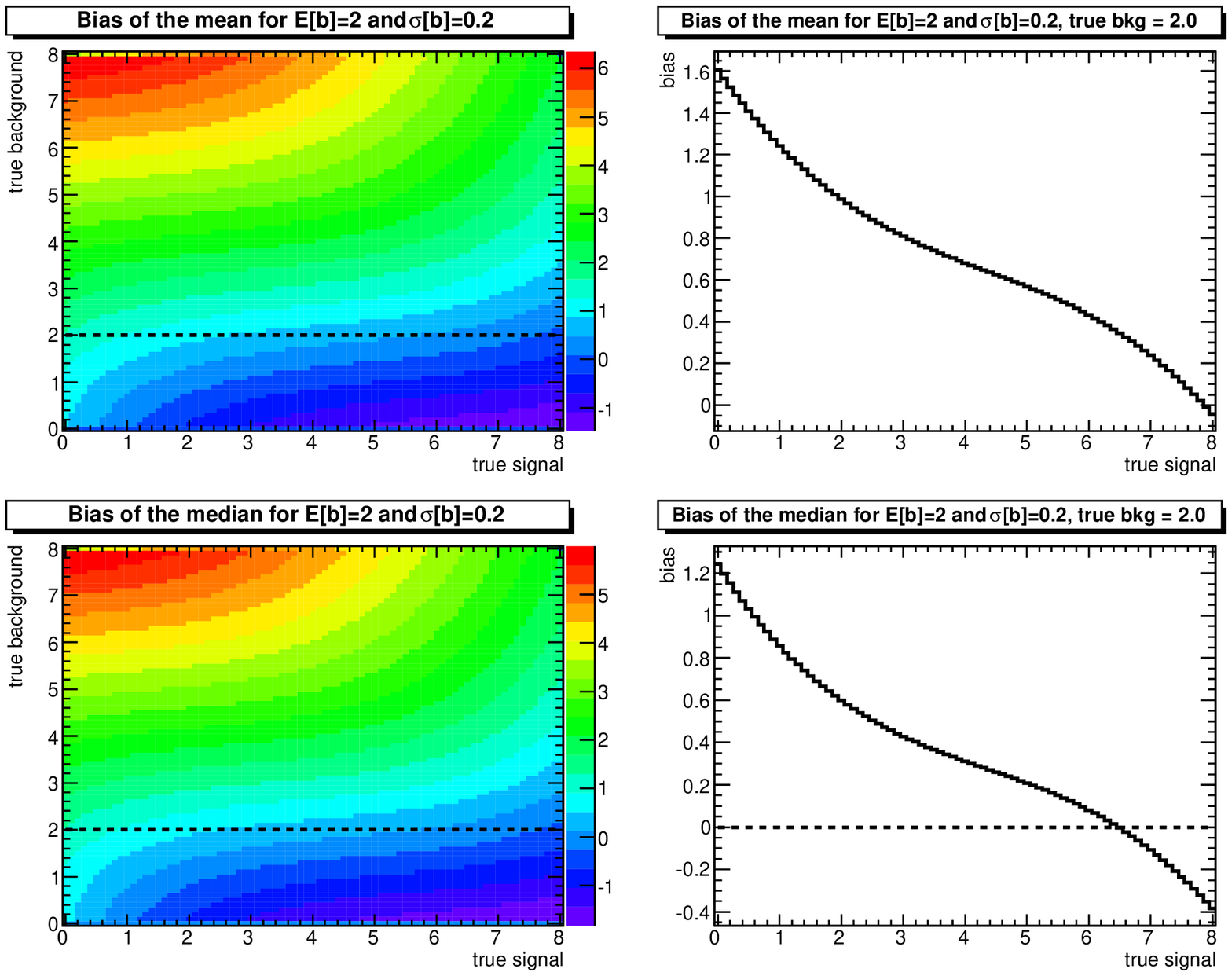}
 \caption{Bias of the posterior mean (top) and median (bottom) as a
   function of the true signal and background (left plots) and of the
   true signal alone for the case in which the true background
   coincides with the prior expectation (right plots).  The posterior
   corresponds to a prior background expectation of 2 counts with 10\%
   relative uncertainty.}
 \label{fig-biasE-biasM-Eb2-Sb10}
\end{figure*}

\begin{figure*}[t!]
 \centering
 \includegraphics[width=\textwidth]{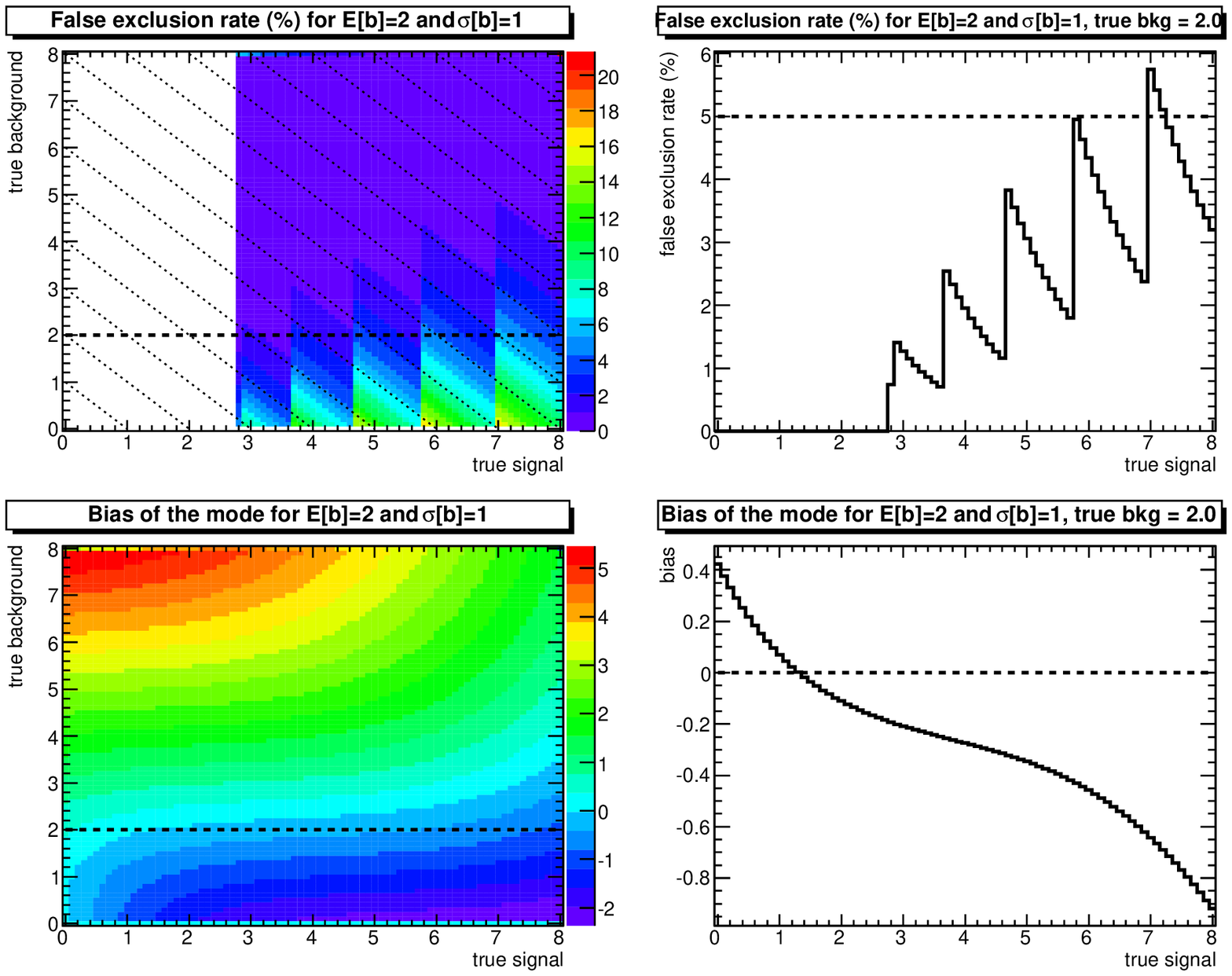}
 \caption{Top: false exclusion rate as a function of the true signal
   and background (top-left) and of the true signal alone for the case
   in which the true background coincides with the prior expectation
   (top-right).  Bottom: bias of the posterior mode as a function of
   the true signal and background (bottom-left) and of the true signal
   alone for the case in which the true background coincides with the
   prior expectation (bottom-right).  The posterior corresponds to a
   prior background expectation of 2 counts with 50\% relative
   uncertainty.}
 \label{fig-FER-biasP-Eb2-Sb50}
\end{figure*}

\begin{figure*}[t!]
 \centering
 \includegraphics[width=\textwidth]{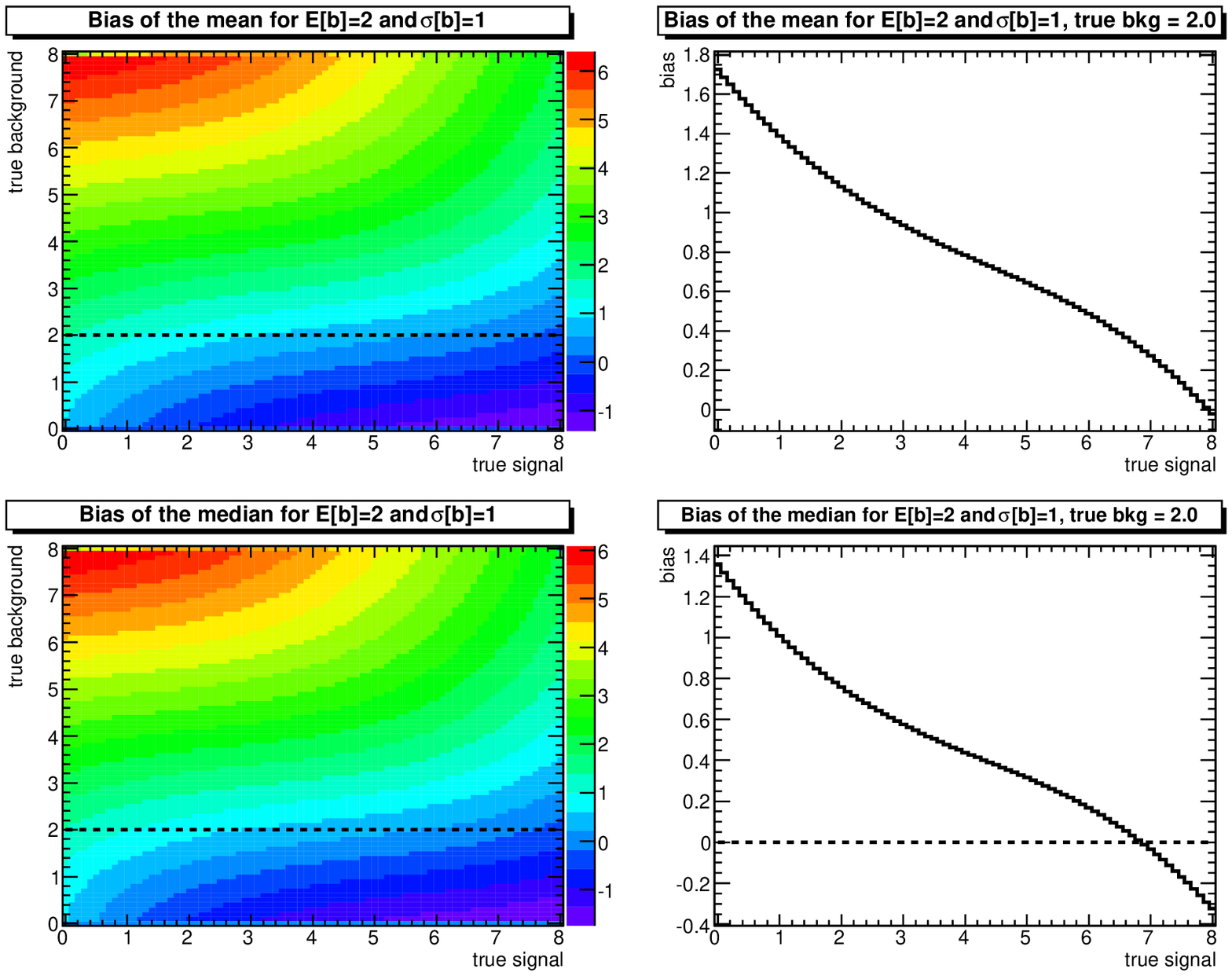}
 \caption{Bias of the posterior mean (top) and median (bottom) as a
   function of the true signal and background (left plots) and of the
   true signal alone for the case in which the true background
   coincides with the prior expectation (right plots).  The posterior
   corresponds to a prior background expectation of 2 counts with 50\%
   relative uncertainty.}
 \label{fig-biasE-biasM-Eb2-Sb50}
\end{figure*}

\begin{figure*}[t!]
 \centering
 \includegraphics[width=\textwidth]{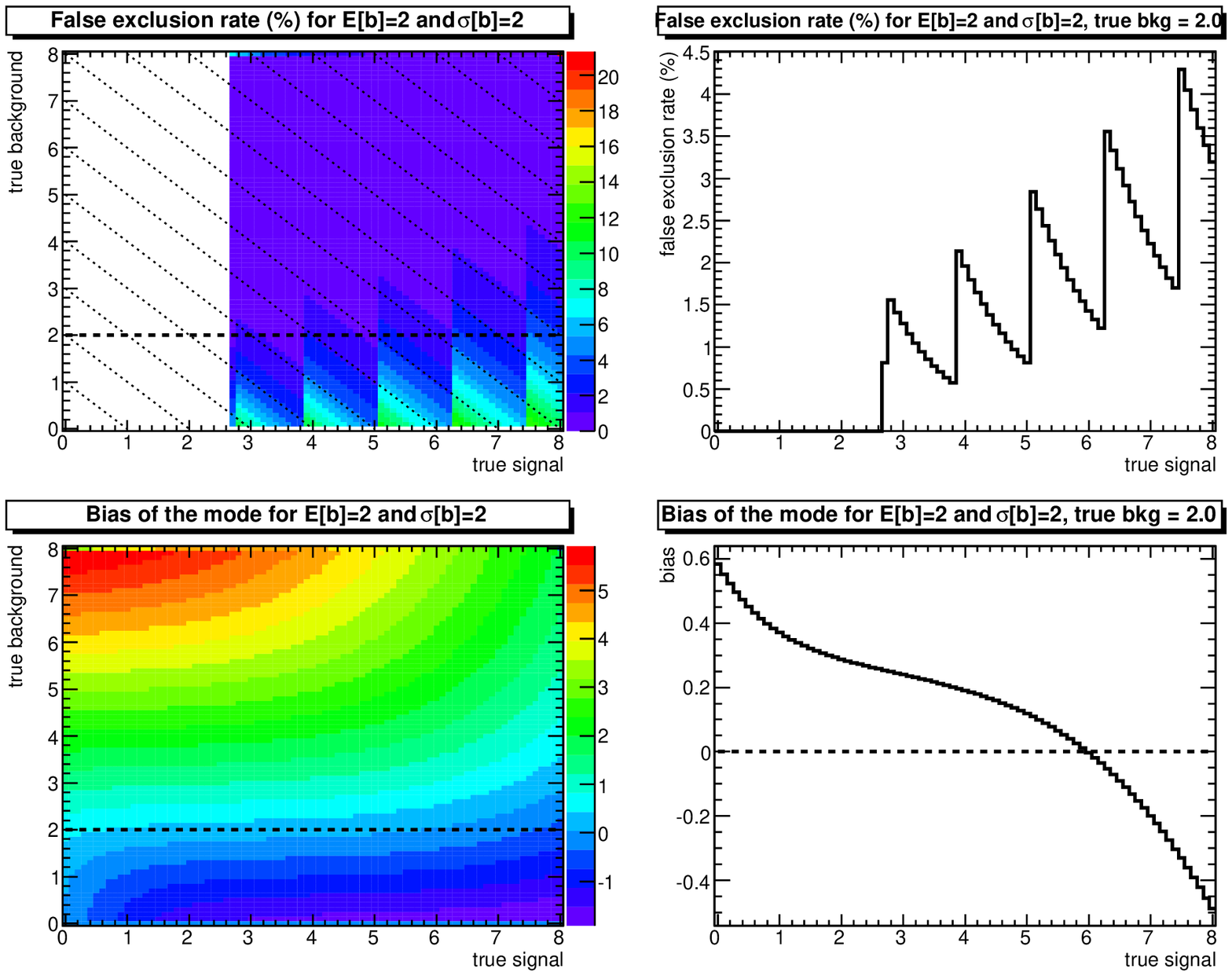}
 \caption{Top: false exclusion rate as a function of the true signal
   and background (top-left) and of the true signal alone for the case
   in which the true background coincides with the prior expectation
   (top-right).  Bottom: bias of the posterior mode as a function of
   the true signal and background (bottom-left) and of the true signal
   alone for the case in which the true background coincides with the
   prior expectation (bottom-right).  The posterior corresponds to a
   prior background expectation of 2 counts with 100\% relative
   uncertainty.}
 \label{fig-FER-biasP-Eb2-Sb100}
\end{figure*}

\begin{figure*}[t!]
 \centering
 \includegraphics[width=\textwidth]{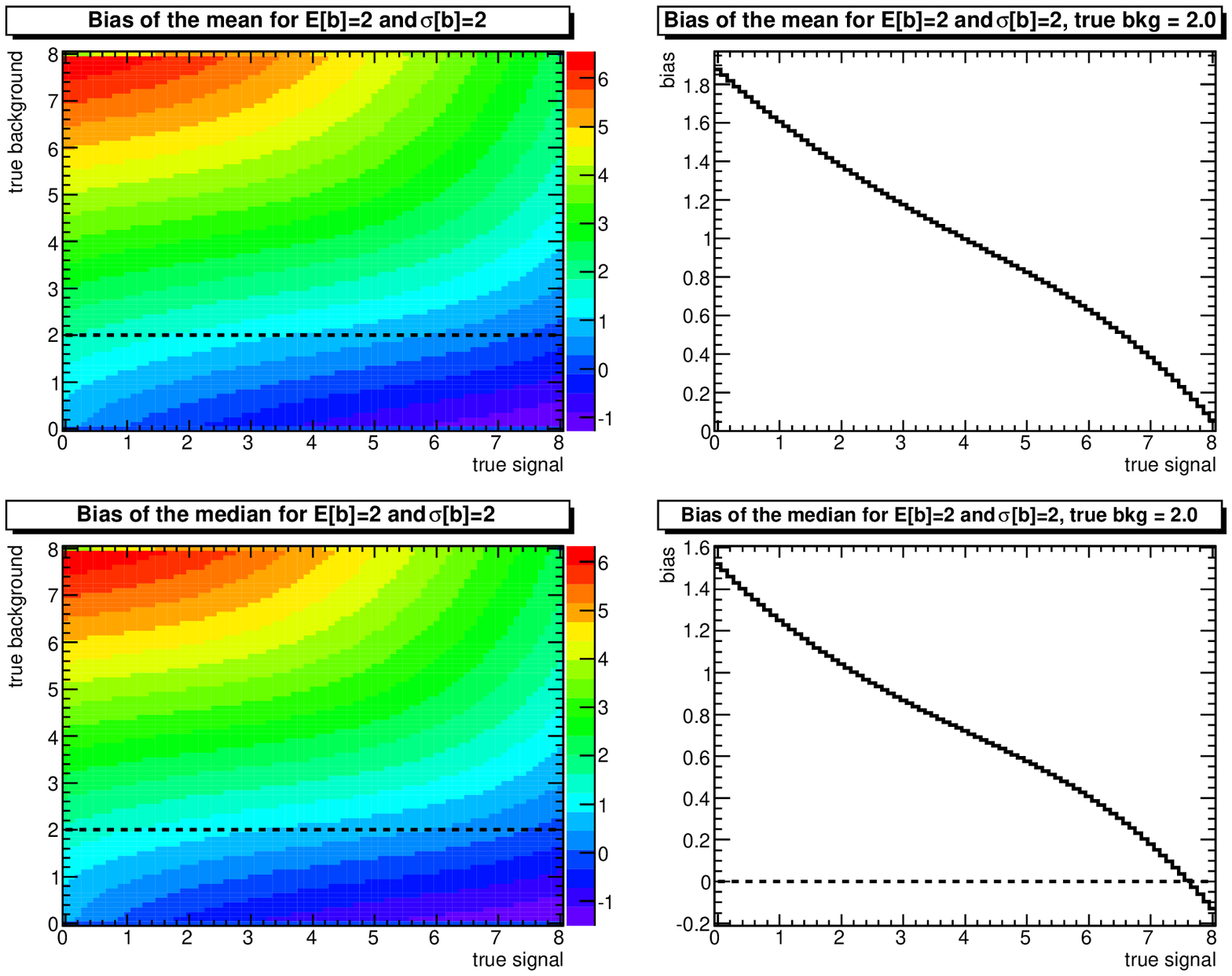}
 \caption{Bias of the posterior mean (top) and median (bottom) as a
   function of the true signal and background (left plots) and of the
   true signal alone for the case in which the true background
   coincides with the prior expectation (right plots).  The posterior
   corresponds to a prior background expectation of 2 counts with 100\%
   relative uncertainty.}
 \label{fig-biasE-biasM-Eb2-Sb100}
\end{figure*}

 Figures \ref{fig-FER-biasP-Eb2-Sb10}, \ref{fig-biasE-biasM-Eb2-Sb10},
 \ref{fig-FER-biasP-Eb2-Sb50}, \ref{fig-biasE-biasM-Eb2-Sb50},
 \ref{fig-FER-biasP-Eb2-Sb100}, and \ref{fig-biasE-biasM-Eb2-Sb100}
 show the false exclusion rate (which is the average fraction of times
 the true signal is above the 95\% upper limit) and the bias of the
 posterior mode, mean and median (which is the difference between the
 estimator and the true signal), as a function of true signal and
 background (left plots) and as a function of the true signal alone,
 for the case in which the true background coincides with the prior
 expectation.  Only the posteriors obtained with a prior background
 expectation $E[b]=2$ counts with relative uncertainties of 10\%, 50\%
 and 100\% are shown.  The empty region in the 2D plots of the false
 exclusion rate corresponds to identically zero exclusion rate: in
 this region the true signal is always smaller than the upper limit.

\begin{sidewaystable}[t!]
\centering
\begin{tabular}{r|rrr|rrr|rrr|rrr}
nObs  &  L95  &  L90  &  L68  &  Mean  &  Median  &  Mode  &  R68  &  R90  &  R95  &  Vari.  &  Skew.  &  Kurt.  \\
\hline                                                  
0  &  0.00  &  0.00  &  0.00  &  0.88  &  0.60  &  0.00  &  1.01  &  2.08  &  2.77  &  0.80  &  2.11  &  6.76  \\
1  &  0.00  &  0.00  &  0.00  &  1.16  &  0.83  &  0.00  &  1.36  &  2.68  &  3.48  &  1.26  &  1.88  &  5.19  \\
2  &  0.00  &  0.00  &  0.00  &  1.56  &  1.18  &  0.00  &  1.86  &  3.46  &  4.38  &  1.94  &  1.62  &  3.73  \\
3  &  0.09  &  0.17  &  0.00  &  2.10  &  1.71  &  0.52  &  2.55  &  5.44  &  6.44  &  2.87  &  1.36  &  2.57  \\
4  &  0.16  &  0.31  &  0.87  &  2.80  &  2.41  &  1.51  &  4.73  &  6.64  &  7.70  &  4.00  &  1.14  &  1.75  \\
5  &  0.34  &  0.59  &  1.39  &  3.63  &  3.26  &  2.51  &  5.85  &  7.91  &  9.04  &  5.21  &  0.96  &  1.24  \\
6  &  0.67  &  1.05  &  2.07  &  4.55  &  4.20  &  3.51  &  7.01  &  9.21  &  10.40  &  6.38  &  0.83  &  0.96  \\
7  &  1.17  &  1.65  &  2.83  &  5.51  &  5.18  &  4.51  &  8.19  &  10.51  &  11.76  &  7.48  &  0.74  &  0.80  \\
8  &  1.77  &  2.32  &  3.64  &  6.50  &  6.17  &  5.51  &  9.36  &  11.80  &  13.10  &  8.52  &  0.69  &  0.70  \\
9  &  2.43  &  3.04  &  4.46  &  7.50  &  7.17  &  6.50  &  10.53  &  13.08  &  14.43  &  9.54  &  0.65  &  0.63  \\
10  &  3.12  &  3.78  &  5.30  &  8.50  &  8.17  &  7.50  &  11.69  &  14.34  &  15.74  &  10.54  &  0.61  &  0.57  \\
11  &  3.82  &  4.53  &  6.15  &  9.50  &  9.16  &  8.50  &  12.84  &  15.59  &  17.04  &  11.54  &  0.59  &  0.52  \\
12  &  4.54  &  5.29  &  7.00  &  10.50  &  10.16  &  9.50  &  13.99  &  16.83  &  18.33  &  12.54  &  0.56  &  0.48  \\
13  &  5.27  &  6.06  &  7.86  &  11.50  &  11.16  &  10.50  &  15.13  &  18.06  &  19.60  &  13.54  &  0.54  &  0.44  \\
14  &  6.00  &  6.84  &  8.73  &  12.50  &  12.16  &  11.50  &  16.27  &  19.28  &  20.87  &  14.54  &  0.52  &  0.41  \\
15  &  6.75  &  7.62  &  9.60  &  13.50  &  13.16  &  12.50  &  17.40  &  20.50  &  22.12  &  15.54  &  0.51  &  0.39  \\
\end{tabular}\caption{Posterior summary for $E[b] = 2 \pm 0.2$ as a function of the number of observed events.}                                                  
\label{tab-post-summ-bkg-2-0.2}
\end{sidewaystable}

\begin{sidewaystable}[t!]
\centering
\begin{tabular}{r|rrr|rrr|rrr|rrr}
nObs  &  L95  &  L90  &  L68  &  Mean  &  Median  &  Mode  &  R68  &  R90  &  R95  &  Vari.  &  Skew.  &  Kurt.  \\
\hline                                                  
0  &  0.00  &  0.00  &  0.00  &  0.88  &  0.60  &  0.00  &  1.01  &  2.08  &  2.78  &  0.81  &  2.10  &  6.73  \\
1  &  0.00  &  0.00  &  0.00  &  1.17  &  0.84  &  0.00  &  1.38  &  2.71  &  3.51  &  1.28  &  1.86  &  5.09  \\
2  &  0.00  &  0.00  &  0.00  &  1.59  &  1.21  &  0.00  &  1.90  &  3.50  &  4.42  &  1.98  &  1.60  &  3.64  \\
3  &  0.09  &  0.17  &  0.53  &  2.13  &  1.74  &  0.57  &  3.75  &  5.50  &  6.49  &  2.92  &  1.35  &  2.50  \\
4  &  0.17  &  0.32  &  0.88  &  2.83  &  2.44  &  1.55  &  4.77  &  6.69  &  7.76  &  4.06  &  1.13  &  1.71  \\
5  &  0.33  &  0.59  &  1.39  &  3.65  &  3.28  &  2.54  &  5.88  &  7.95  &  9.09  &  5.28  &  0.95  &  1.21  \\
6  &  0.64  &  1.02  &  2.06  &  4.56  &  4.21  &  3.53  &  7.04  &  9.25  &  10.44  &  6.47  &  0.82  &  0.93  \\
7  &  1.12  &  1.61  &  2.81  &  5.52  &  5.18  &  4.53  &  8.21  &  10.54  &  11.79  &  7.59  &  0.73  &  0.77  \\
8  &  1.71  &  2.28  &  3.61  &  6.50  &  6.17  &  5.52  &  9.39  &  11.83  &  13.14  &  8.65  &  0.67  &  0.68  \\
9  &  2.37  &  2.99  &  4.44  &  7.50  &  7.17  &  6.52  &  10.55  &  13.10  &  14.46  &  9.67  &  0.63  &  0.61  \\
10  &  3.06  &  3.73  &  5.28  &  8.50  &  8.17  &  7.52  &  11.71  &  14.37  &  15.77  &  10.68  &  0.60  &  0.55  \\
11  &  3.77  &  4.49  &  6.13  &  9.50  &  9.17  &  8.52  &  12.86  &  15.62  &  17.07  &  11.68  &  0.58  &  0.51  \\
12  &  4.49  &  5.25  &  6.98  &  10.50  &  10.17  &  9.51  &  14.01  &  16.85  &  18.36  &  12.68  &  0.55  &  0.47  \\
13  &  5.22  &  6.02  &  7.84  &  11.50  &  11.17  &  10.51  &  15.15  &  18.08  &  19.63  &  13.68  &  0.53  &  0.43  \\
14  &  5.96  &  6.80  &  8.71  &  12.50  &  12.17  &  11.51  &  16.28  &  19.30  &  20.89  &  14.67  &  0.52  &  0.40  \\
15  &  6.71  &  7.59  &  9.58  &  13.50  &  13.17  &  12.51  &  17.41  &  20.52  &  22.14  &  15.67  &  0.50  &  0.38  \\
\end{tabular}\caption{Posterior summary for $E[b] = 2 \pm 0.4$ as a function of the number of observed events.}                                                  
\label{tab-post-summ-bkg-2-0.4}
\end{sidewaystable}

\begin{sidewaystable}[t!]
\centering
\begin{tabular}{r|rrr|rrr|rrr|rrr}
nObs  &  L95  &  L90  &  L68  &  Mean  &  Median  &  Mode  &  R68  &  R90  &  R95  &  Vari.  &  Skew.  &  Kurt.  \\
\hline                                                  
0  &  0.00  &  0.00  &  0.00  &  0.89  &  0.61  &  0.00  &  1.02  &  2.10  &  2.80  &  0.82  &  2.10  &  6.68  \\
1  &  0.00  &  0.00  &  0.00  &  1.26  &  0.92  &  0.00  &  1.48  &  2.86  &  3.67  &  1.39  &  1.78  &  4.64  \\
2  &  0.00  &  0.00  &  0.00  &  1.73  &  1.35  &  0.00  &  2.08  &  3.74  &  4.68  &  2.18  &  1.51  &  3.20  \\
3  &  0.10  &  0.20  &  0.61  &  2.31  &  1.92  &  0.82  &  4.02  &  5.79  &  6.80  &  3.19  &  1.27  &  2.18  \\
4  &  0.17  &  0.33  &  0.95  &  3.00  &  2.62  &  1.77  &  5.03  &  6.98  &  8.06  &  4.38  &  1.06  &  1.49  \\
5  &  0.30  &  0.56  &  1.42  &  3.78  &  3.43  &  2.73  &  6.11  &  8.22  &  9.37  &  5.67  &  0.90  &  1.05  \\
6  &  0.52  &  0.91  &  2.02  &  4.65  &  4.32  &  3.70  &  7.24  &  9.48  &  10.69  &  6.97  &  0.77  &  0.78  \\
7  &  0.87  &  1.39  &  2.72  &  5.57  &  5.26  &  4.68  &  8.38  &  10.75  &  12.02  &  8.22  &  0.67  &  0.63  \\
8  &  1.36  &  2.00  &  3.49  &  6.52  &  6.22  &  5.66  &  9.53  &  12.02  &  13.34  &  9.41  &  0.60  &  0.54  \\
9  &  1.96  &  2.68  &  4.31  &  7.50  &  7.21  &  6.64  &  10.69  &  13.28  &  14.66  &  10.52  &  0.55  &  0.49  \\
10  &  2.63  &  3.42  &  5.14  &  8.49  &  8.20  &  7.63  &  11.84  &  14.53  &  15.96  &  11.59  &  0.52  &  0.46  \\
11  &  3.34  &  4.18  &  5.99  &  9.49  &  9.20  &  8.62  &  12.98  &  15.78  &  17.25  &  12.62  &  0.49  &  0.43  \\
12  &  4.08  &  4.96  &  6.85  &  10.49  &  10.19  &  9.61  &  14.12  &  17.01  &  18.52  &  13.63  &  0.48  &  0.41  \\
13  &  4.83  &  5.74  &  7.72  &  11.49  &  11.19  &  10.60  &  15.26  &  18.23  &  19.79  &  14.63  &  0.46  &  0.38  \\
14  &  5.60  &  6.54  &  8.59  &  12.49  &  12.19  &  11.60  &  16.39  &  19.45  &  21.05  &  15.62  &  0.45  &  0.36  \\
15  &  6.36  &  7.34  &  9.47  &  13.49  &  13.19  &  12.59  &  17.52  &  20.66  &  22.30  &  16.62  &  0.44  &  0.34  \\
\end{tabular}\caption{Posterior summary for $E[b] = 2 \pm 1$ as a function of the number of observed events.}                                                  
\label{tab-post-summ-bkg-2-1}
\end{sidewaystable}

\begin{sidewaystable}[t!]
\centering
\begin{tabular}{r|rrr|rrr|rrr|rrr}
nObs  &  L95  &  L90  &  L68  &  Mean  &  Median  &  Mode  &  R68  &  R90  &  R95  &  Vari.  &  Skew.  &  Kurt.  \\
\hline
0  &  0.00  &  0.00  &  0.00  &  0.84  &  0.56  &  0.00  &  0.96  &  2.03  &  2.74  &  0.79  &  2.17  &  7.17  \\
1  &  0.00  &  0.00  &  0.00  &  1.36  &  1.03  &  0.00  &  1.62  &  3.04  &  3.86  &  1.52  &  1.69  &  4.18  \\
2  &  0.00  &  0.00  &  0.00  &  1.95  &  1.59  &  0.00  &  2.37  &  4.10  &  5.05  &  2.48  &  1.37  &  2.63  \\
3  &  0.12  &  0.24  &  0.74  &  2.60  &  2.24  &  1.40  &  4.45  &  6.26  &  7.28  &  3.64  &  1.13  &  1.72  \\
4  &  0.18  &  0.36  &  1.08  &  3.31  &  2.97  &  2.32  &  5.50  &  7.49  &  8.59  &  4.96  &  0.94  &  1.14  \\
5  &  0.27  &  0.53  &  1.51  &  4.08  &  3.77  &  3.26  &  6.58  &  8.73  &  9.90  &  6.41  &  0.78  &  0.76  \\
6  &  0.39  &  0.76  &  2.03  &  4.90  &  4.62  &  4.20  &  7.68  &  9.98  &  11.21  &  7.92  &  0.65  &  0.52  \\
7  &  0.56  &  1.07  &  2.64  &  5.76  &  5.52  &  5.16  &  8.79  &  11.22  &  12.51  &  9.46  &  0.55  &  0.37  \\
8  &  0.79  &  1.46  &  3.32  &  6.65  &  6.44  &  6.12  &  9.92  &  12.46  &  13.81  &  10.99  &  0.46  &  0.29  \\
9  &  1.11  &  1.95  &  4.05  &  7.58  &  7.39  &  7.08  &  11.04  &  13.70  &  15.10  &  12.47  &  0.38  &  0.25  \\
10  &  1.52  &  2.54  &  4.84  &  8.52  &  8.35  &  8.05  &  12.17  &  14.94  &  16.39  &  13.89  &  0.33  &  0.23  \\
11  &  2.02  &  3.20  &  5.66  &  9.49  &  9.33  &  9.02  &  13.30  &  16.16  &  17.66  &  15.24  &  0.28  &  0.24  \\
12  &  2.62  &  3.92  &  6.50  &  10.46  &  10.31  &  10.00  &  14.43  &  17.39  &  18.93  &  16.53  &  0.24  &  0.26  \\
13  &  3.29  &  4.69  &  7.36  &  11.45  &  11.30  &  10.98  &  15.56  &  18.60  &  20.19  &  17.75  &  0.22  &  0.28  \\
14  &  4.03  &  5.49  &  8.23  &  12.44  &  12.29  &  11.96  &  16.68  &  19.81  &  21.44  &  18.92  &  0.20  &  0.30  \\
15  &  4.81  &  6.31  &  9.11  &  13.44  &  13.28  &  12.94  &  17.80  &  21.01  &  22.68  &  20.04  &  0.19  &  0.32  \\
\end{tabular}
\caption{Posterior summary for $E[b] = 2 \pm 2$ as a function of the
  number of observed events.}
\label{tab-post-summ-bkg-2-2}
\end{sidewaystable}

\clearpage

\begin{figure*}[t!]
 \centering
 \includegraphics[width=\textwidth]{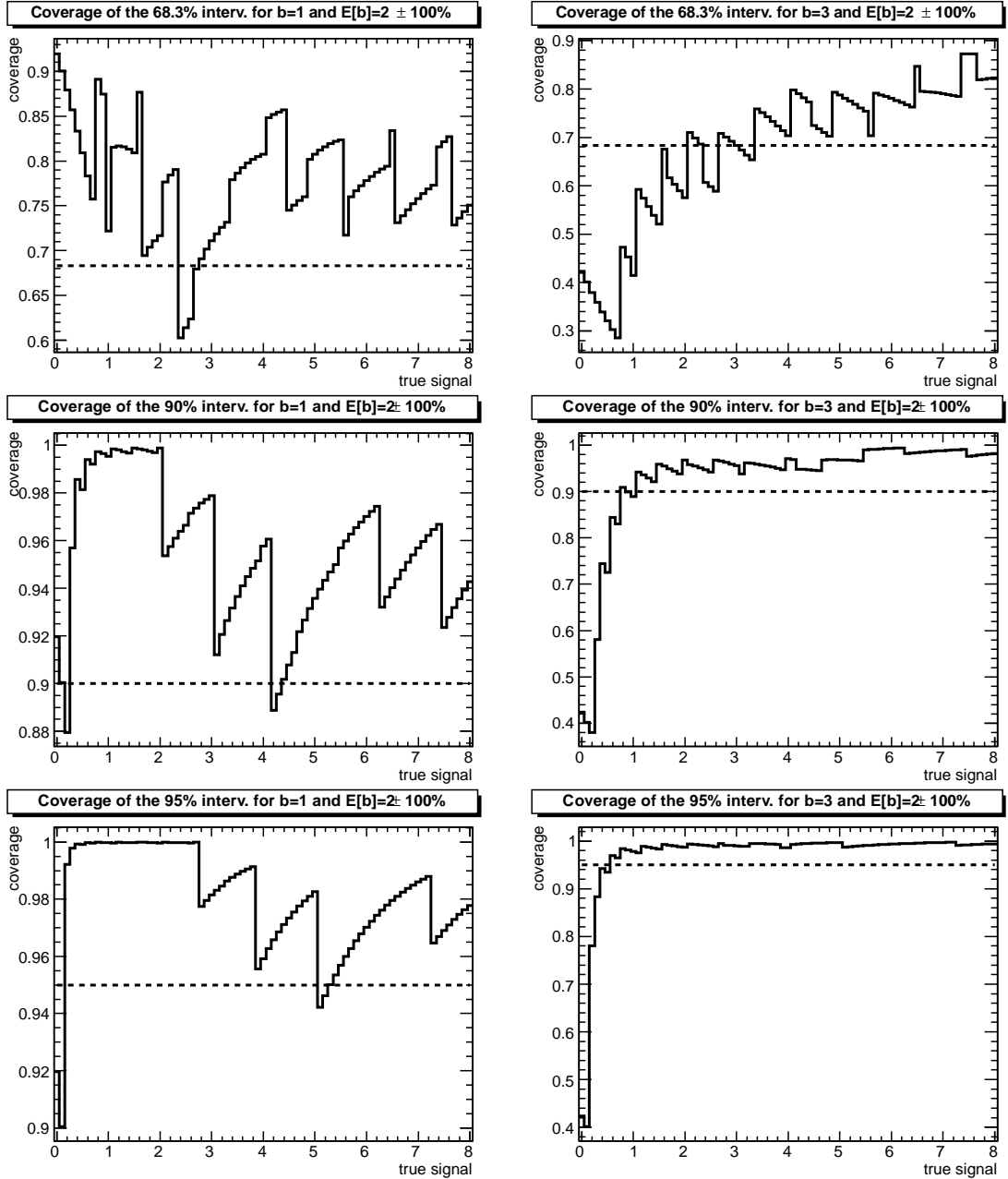}
 \caption{Coverage of the 95\% (top), 90\% (middle), 68.3\% (bottom)
 posterior credible intervals for $E[b] = 2 \pm 2$ as a function of
 the true signal, when the true background is half of the background
 expectation (left) or 50\% bigger (right).}
 \label{fig-coverage-Eb2-2-b1e3}
\end{figure*}


\end{document}